\pdfoutput=1
\documentclass{aastex62}
\usepackage{natbib}
\usepackage{graphicx}

\received{January 1, 0000}
\revised{January 1, 0000}
\accepted{January 1, 0000}
\submitjournal{ApJ}

\shorttitle{ML in IIL/P SNe}
\shortauthors{Chen et al.}

\begin{document}
\title{A New Method to Classify Type IIP/IIL Supernovae Based on their Spectra }
\correspondingauthor{Xuewen Liu}
\email{liuxuew@scu.edu.cn}

\author{Xingzhuo Chen}
\affiliation{College of Physical Science and Technology,\\
Sichuan University, Chengdu, 610064,\\
People Republic of China}

\author{Shihao Kou}
\affiliation{College of Physical Science and Technology,\\
Sichuan University, Chengdu, 610064,\\
People Republic of China}

\author{Xuewen Liu}
\affiliation{College of Physical Science and Technology,\\
Sichuan University, Chengdu, 610064,\\
People Republic of China}

\begin{abstract}
  Type IIP and type IIL supernovae (SNe) are defined on their light curves, but the
  spectrum criteria in distinguishing these two type SNe remains unclear. We propose a new
  classification method. Firstly, we subtract the principal components of different wavelength bands in the spectra based on Functional Principal Components Analysis (FPCA) method. 
Then, we use Support Vector Machine (SVM) and Artificial Neural Network (ANN) to classify these two types of SNe. 
The best F1-Score of our classifier is 0.881, and we found that solely using $H_\alpha$ line at 6150-6800 \AA \ for classification can reach a F1-Score up to 0.849. Our result
indicates that the profile of the $H_\alpha$ is the key to distinguish the two type SNe.
\end{abstract}


\section{introduction}
Originated from massive stars($8\sim20M_\sun$), Hydrogen Rich Core Collapse (CC) Supernovae(SNe) also known as type II SNe is one kind of SNe with conspicuous hydrogen spectral line at 6335 \AA \ \citep{gyobservational}. 
Some type II SNe are generally divided into 4 subtypes: IIP, IIL, IIb and IIn. 
Moreover, some SNe with peculiar spectroscopic or light curve features are divided into type II-pec SNe, such as SN1987A \citep{1987aphysics}. 
Helium spectral lines at the wavelength 5876,6678,7065 \AA \  are observable in type IIb, while Type IIn  spectra usually have a narrow $H_\alpha$ line at 6563 \AA. 
As for Type IIP/IIL SNe, their definitions are based on the light curves. 

The lumiosity evolution of SNe are generally dominated by 4 mechanisms\citep{hrichccsne}: shock breakout, shock cooling \& ejecta recombination, radioactive decay and circumstllar material (CSM) interaction. 
Type IIP SNe are characteristic of its plateau-like lightcurves, which fast rises to the peak ($\sim$ 15 days) after explosion and follows a plateau-phase ($\sim$90 days) which is powered from the cooling of ejecta.
After the `plateau phase' the magnitude will linearly decline ($>$1.4 Mag/100 day) with the propulsion of the nikel decay $^{56}{\rm Ni}\rightarrow{^{56}\rm Co}\rightarrow{^{56}\rm Fe}$. 
In contrast, Type IIL SNe' cooling phase is much shorter than Type IIP (typically less than 10 days) and its luminosity linearly decrease ($\sim$0.3Mag/15day)\citep{gyobservational} after reaching maximum. 
Notably, Type IIb SNe has the most various lightcurve and some of them even shows a double-peak light curves, the first of which is originated from the fast cooling effect after the explosion, while the second of which links to the high-energy photon escaping the relatively small-mass envelope\citep{hrichccsne,liweidongluminosity,sn1993j}.

Many studies indicates the progenitor of Type IIP SNe is Red Super Giants (RSG) with mass range of 8.5-16.5$M_\sun$ \citep{rsgiip}, while Type IIb SNe are generally originated from Yellow Super Giants (YSG)\citep{2IILstar}. 
Moreover, some studies suggest Type IIn SNe' progenitors are Luminous Blue Variables because the interaction between the core and Circumstellar Materials (CSM)\citep{initialiin}. 
Nonetheless, the progenitor of Type IIL SNe is on debate and only one observational evidence on SN2009kr suggests RSG or YSG could be progenitor\citep{IILprogenitor,sn2009krysg}. 

In the former studies on Type IIP/IIL SNe, the absorption-emission ratio of $H_\alpha$ P-Cygni profile in Type IIL SNe is relatively smaller than Type IIP's, which is possibly because of Type IIP/IIL SNe's envelope mass and density gradient differences after the peak magnitude\citep{absorpemit1,absorpemit2}. 
Moreover, in a study on $H_\alpha$ and O(I)7774 equivalent width(EW), the spectra of Type IIP SNe have a smaller ratio $EW_{OI7774}/EW_{H_\alpha}$\citep{faranIIPL2}. 
Additionally, at the peak luminosity, Type IIL SNe' absolute magnitude is $-17.44\pm0.22$ while Type IIP' is $-15.66\pm0.16$\citep{liweidongluminosity}. 
However, unlike the classification in Type I SNe\citep{sfwia}, scant spectroscopic criteria are presented to determine Type IIP/IIL SNe. 

Recently, with the advance of machine-learning algorithm and applications, new spectroscopic and photometric classification methods are introduced. 
Several studies have been conducted in photometric classification and instant detection of SNe and other transients via machine learning algorithms\citep{mlsdss,photoml2,photoml3}. 
With Principal Component Analysis (PCA) and Functional Principal Analysis (FPCA) algorithm, the light curves of SNe are able to described into fewer parameters\citep{typeiicluster,wlffpca}. 
Furthermore, combining PCA with other machine learning algorithms, i.e. Self Organizing Maps and K-means, the spectral features in different subtypes of Type Ia SNe has been studied\citep{pcaia}. 

In this study, we applied Functional Principal Component Analysis (FPCA) on the different wave bands of Type IIP/IIL SNe' spectra, then use Support Vector Machine (SVM) and Artificial Neural Network (ANN) to classify the types of SNe from the principal components (FPCA scores). 
In \S~\ref{sec:method}, we will introduce the data source and the algorithms (FPCA, SVM and ANN) used in the paper. 
In \S~\ref{sec:result}, we will discuss the performance of our classifiers. 
The prospective and the summary will be presented in \S~\ref{sec:summary}. All the codes used in the analysis are uploaded onto \url{https://github.com/GeronimoChen/IIL-IIP-SNe}. 

\section{method}\label{sec:method}

\subsection{Pre-processing the data}\label{sec:data}

\begin{figure}
\plottwo{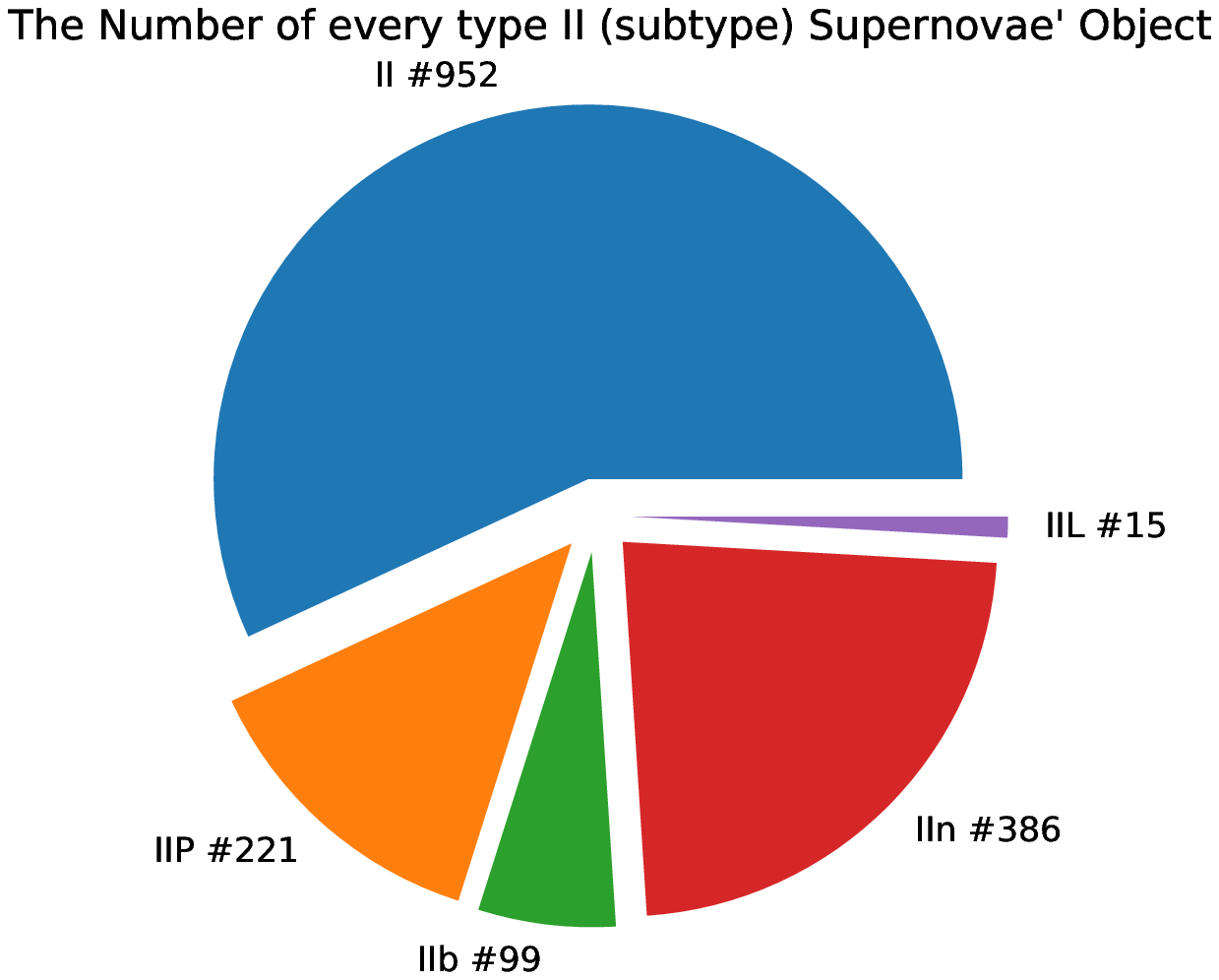}{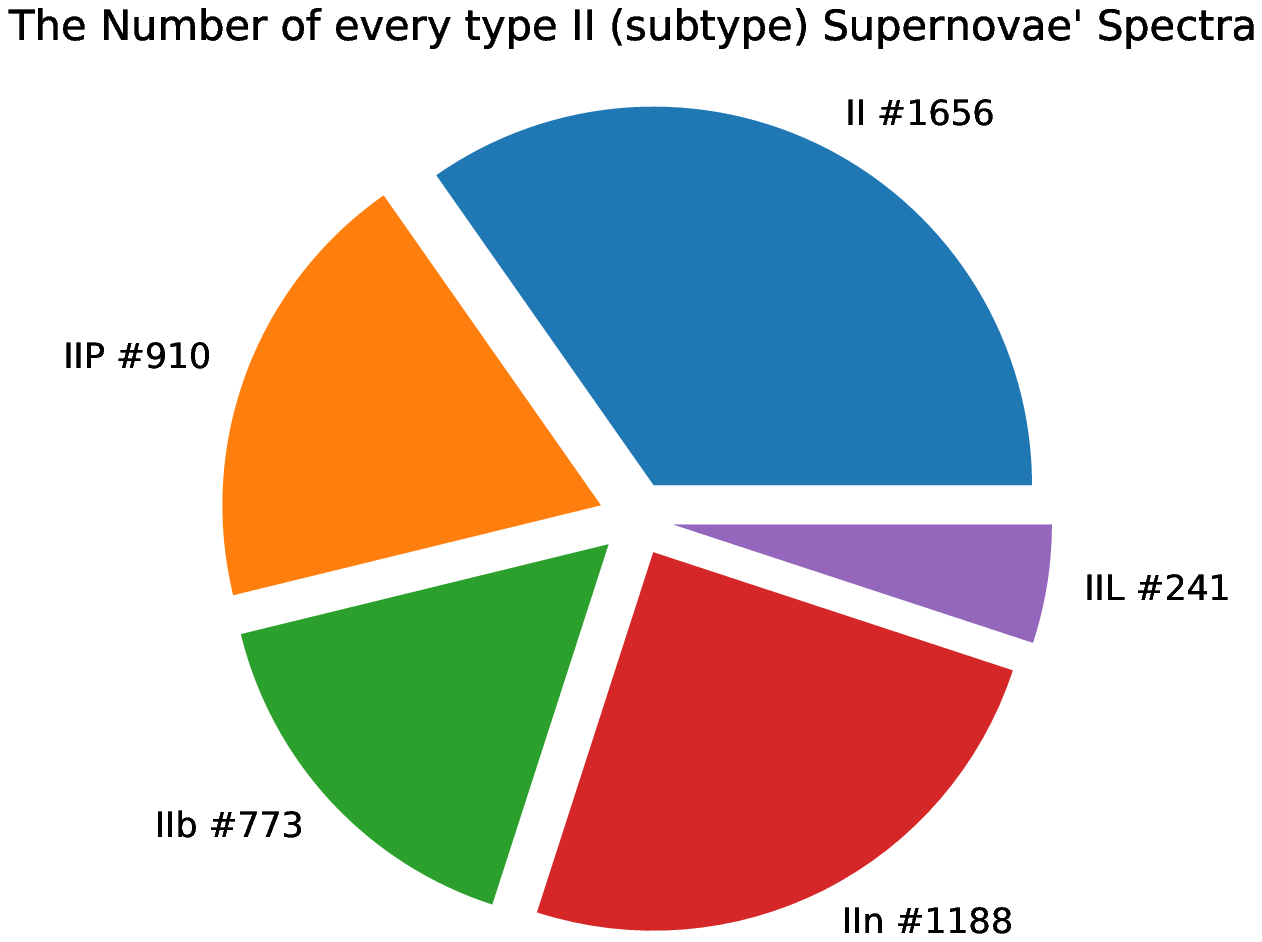}
\caption{The pie chart of Type II/IIP/IIL/IIb/IIn objects and spectra. The number of objects/data are marked with the type.}\label{fig:data}
\end{figure}

All the spectra of Type II SNe (/IIP/IIL/IIb/IIn) are downloaded from WISeREP\citep{wiserep} for our analysis.
Because there are only Type IIL in the database, we tried to find 6 more Type IIL SNe from other literatures to expand the dataset, which is shown in Table.\ref{tab:moreiil}. 
As the unclassified Type IIL has been concluded, the dispersion of different types' SNe object and spectra are shown in Fig.\ref{fig:data}.

\begin{deluxetable}{c|ccc}[ht!]
\tablecaption{More Type IIL SNe from Other Literature}\label{tab:moreiil}
\tablehead{
\colhead{Supernovae Name} & \colhead{Number of Spectra} & \colhead{Type in WISeREP} & \colhead{Reference}
}
\startdata
SN1980K & 6 & II & \citep{1980k} \\
SN1979C & 4 & II & \citep{1979c} \\
SN2013hj & 5 & II & \citep{2013hj} \\
SN2013by & 18 & II & \citep{2013by} \\
SN2013ej & 112 & II & \citep{2013ej} \\
SN2009kr & 2 & IIn & \citep{2IILstar} \\
\enddata

\end{deluxetable}

\begin{figure}
\plotone{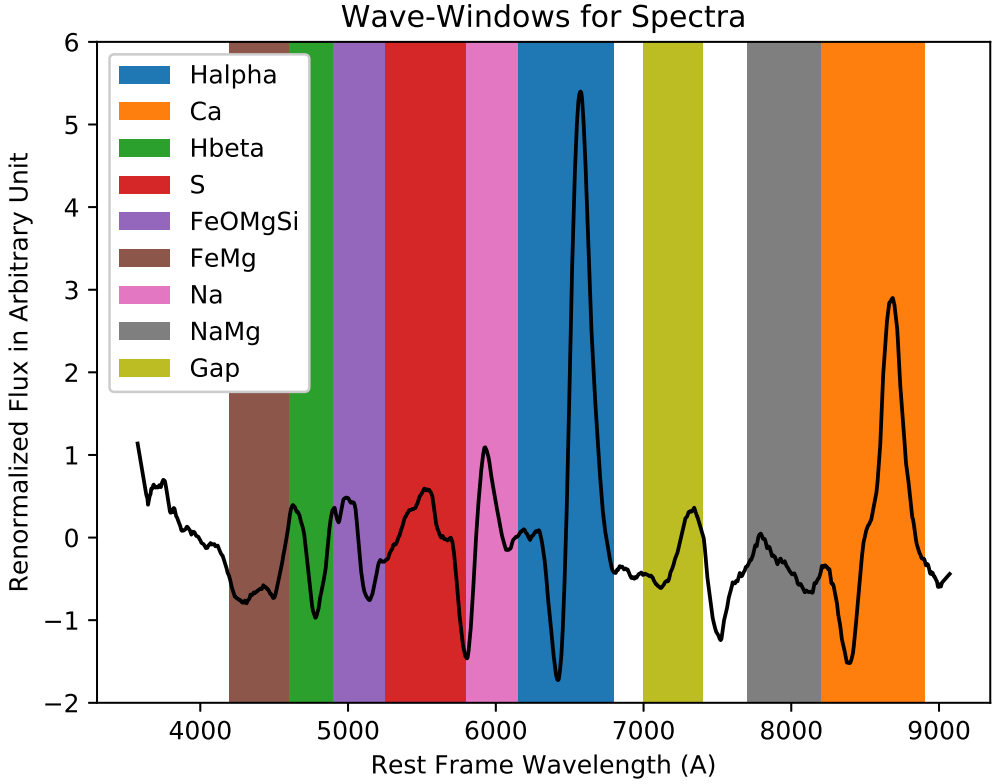}
\caption{The wave-windows chosen for analysis. The range and the element spectra composition are listed in Table~\ref{tab:wavewindow}.} \label{fig:ill}
\end{figure}

The pre-processing of the spectra are discussed in the Section~\ref{sec:methods}, which is constitute of removing redshifts, Savitz-Golay filter\citep{sgfilter}, removing continuum spectra. 
Then, the as-prepared spectra are divided into 9 small wavelength windows and 2 long wavelength windows, as is shown in Fig.\ref{fig:ill}. 
Refering to peer literatures\citep{asassn14dq,2006gy,metallicity,llhostgalaxy,lowmetallicity,sn2016esw,1979c2,sn2004et}, we have concluded the major spectral lines in SN identification into the Table~\ref{tab:wavewindow}. 
Because of the conspicuous hygrogen line in most Type II SNe, we carefully selected the $H_\alpha$ and $H_\beta$ wave-windows to include all the absorption and emission parts for most spectra. 
Most the small wave-windows are denoted with the spectral lines in the spectra, except the wave-window `Gap', which was excluded initially due to the $H_2O$ spectral line at 7165 \AA. 
However, some Type II SNe may embody calcium spectral line (Ca(II)7291,7323) in this reagion\citep{sn2016esw,sn2004et,faranIIPL2}, so this wavelength window is preserved. 
Furthermore, wavelength at 6800-7000 and 7400-7700 \AA \ are excluded from the small wavelength windows due to the telluric spectral line ($O_2\ 6867,\ 7620$). 
Relating to a study in PCA analysis in Type Ia spectra\citep{pcaia}, we kept two long wave-windows here, which are the `visible' wavelength at 4000-7000 \AA \ and 'expand' wavelength at 4000-9000 \AA. 
Because of the equipments limitation and redshift effect, some wave-windows are not fully sampled or even not sampled in a spectrum, which cause an extra data loss in the following training for the classifying algorithm. 
The number available spectra in each wave-windows are also counted in Table~\ref{tab:wavewindow}.

\begin{deluxetable}{c|c|c|c}
\tablecaption{Wave-Window Ranges and Spectral Lines}\label{tab:wavewindow}
\tablehead{
\colhead{Name} & \colhead{Wavelength Range (\AA)} & \colhead{Elements and Spectral Lines} & \colhead{Avaliable Type IIL/IIP spectra}
}
\startdata
FeMg & 4200-4600 & $H_\gamma$4340, $H_\delta$4102,  Ba(II)4554& 170/667\\
$H_\beta$ & 4600-4900 & $H_\beta$4861 & 181/717\\
FeOMgSi & 4900-5250 & Fe(II)4924,5018,5108,5169. O(III)4959,5007. & 176/733\\
S & 5250-5800 &S(II)5454,5433. O(V)5597.& 154/700\\
Na & 5800-6150 & Na(I)5876,5896. He(I)5876. Ba(II)6142. & 168/722\\
$H_\alpha$ & 6150-6800 & $H_\alpha$ 6563. He(I)6678, O(I)6300,6364. N(II)6548. Sc(II)6247& 177/735\\
Gap & 7000-7400 & Fe(II)7155. He(I)7065. O(II)7319,7330. Ca(II)7291,7323& 159/672\\
NaMg & 7700-8200 & Na(I)8183,8195 & 111/491\\
Ca & 8200-8900 & Ca(II)8498,8542,8662. O(I)7774.  & 113/440\\
Visible & 4000-7000 & /& 115/567\\
Expand & 4000-9000 & /& 63/306\\
\enddata
\tablecomments{The names of each wave-window are roughly selected based on the spectral line given in WISeREP, but not all the elements exhibit in Type II SNe. }
\end{deluxetable}

\subsection{FPCA algorithm}

We use Functional Principal Components Analysis (FPCA) for the first process of each wave-window in order to acquire a series principal scores of each spectrum. This algorithm was applied to SNe' light curve analysis before\citep{wlffpca}. 

The FPCA algorithm aims to give a set of orthonormal functions, which could represent a series of trajectory via their linear combination\citep{fpca1,fpca2,fpca3}. 
Considering n SNe spectra, $X_n(\lambda)$, which could be represented in the following equations:

\begin{equation}
X_n(\lambda)=\mu(\lambda)+\sum_{m=1}^\infty \beta_{m,n} \phi_m(\lambda)
\end{equation}

Where $\beta_m$ is the m-th FPCS scores, $\phi_m(\lambda)$ is the m-th basis functions and $\mu(\lambda)$ is the average spectrum of all selected spectra. 
Because the mean function has been subtracted into $\mu(\lambda)$, the average of each order FPCS scores for all spectra are zero. 
Every basis functions preserves orthonormal conditions, and the order of basis functions are sorted by the variance of FPCS socres: 

\begin{equation}
\int_{\lambda_{min}}^{\lambda_{max}}\phi_j(\lambda)\phi_k(\lambda)d\lambda=\delta_{jk}
\end{equation}
\begin{equation}
\forall m: Var(\beta_{m,n=1,2...n_{max}})>Var(\beta_{m+1,n=1,2,...n_{max}})
\end{equation}

As every basis functions are orthonormal, the m-th FPCS score of a single spectrum can be calculated via convolution in the selected wave-window $\lambda_{min}\sim \lambda_{max}$: 

\begin{equation}\label{eq:convolution}
\beta_{m,n}=\int_{\lambda_{min}}^{\lambda_{max}}\phi_m(\lambda)\left(X_n(\lambda)-\mu(\lambda)\right)d\lambda
\end{equation}

In order to preserve the most information of spectra $X(\lambda)$, the FPCS scores of each order should as spacse as possible (the variance is as big as possible), and the selection of basis functions obeys the following equation. 

\begin{equation}
\phi_n(\lambda)=argmax\left(Var\left(\int_{\lambda_{min}}^{\lambda_{max}}\left(X(\lambda)-\mu(\lambda)\right)\phi_n(\lambda)d\lambda \right) \right)
\end{equation}

In this literature, we adopt {\tt\string fpca} package in R language \citep{fpca3} for FPCA analysis, which use a gradient optimize method to acquire the best set of basis functions. 
With the order of basis function increases, the FPCS scores are converge to zero, which gives possible to set a highest order and perserves most informations in the raw spectra $X_n(\lambda)$. 

\begin{figure}[!htb]
\minipage{0.32\textwidth}
  \includegraphics[width=\linewidth]{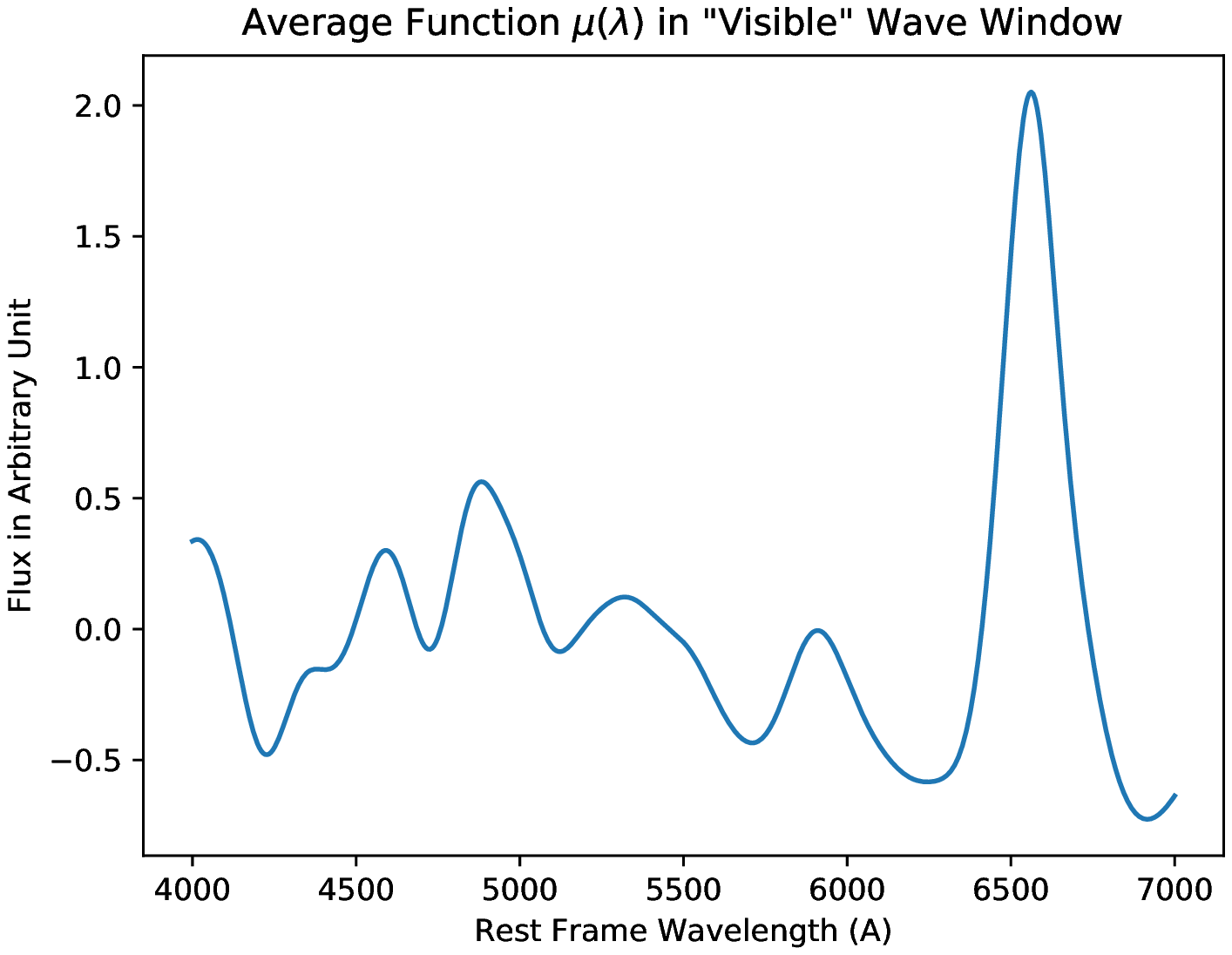}
\endminipage\hfill
\minipage{0.32\textwidth}
  \includegraphics[width=\linewidth]{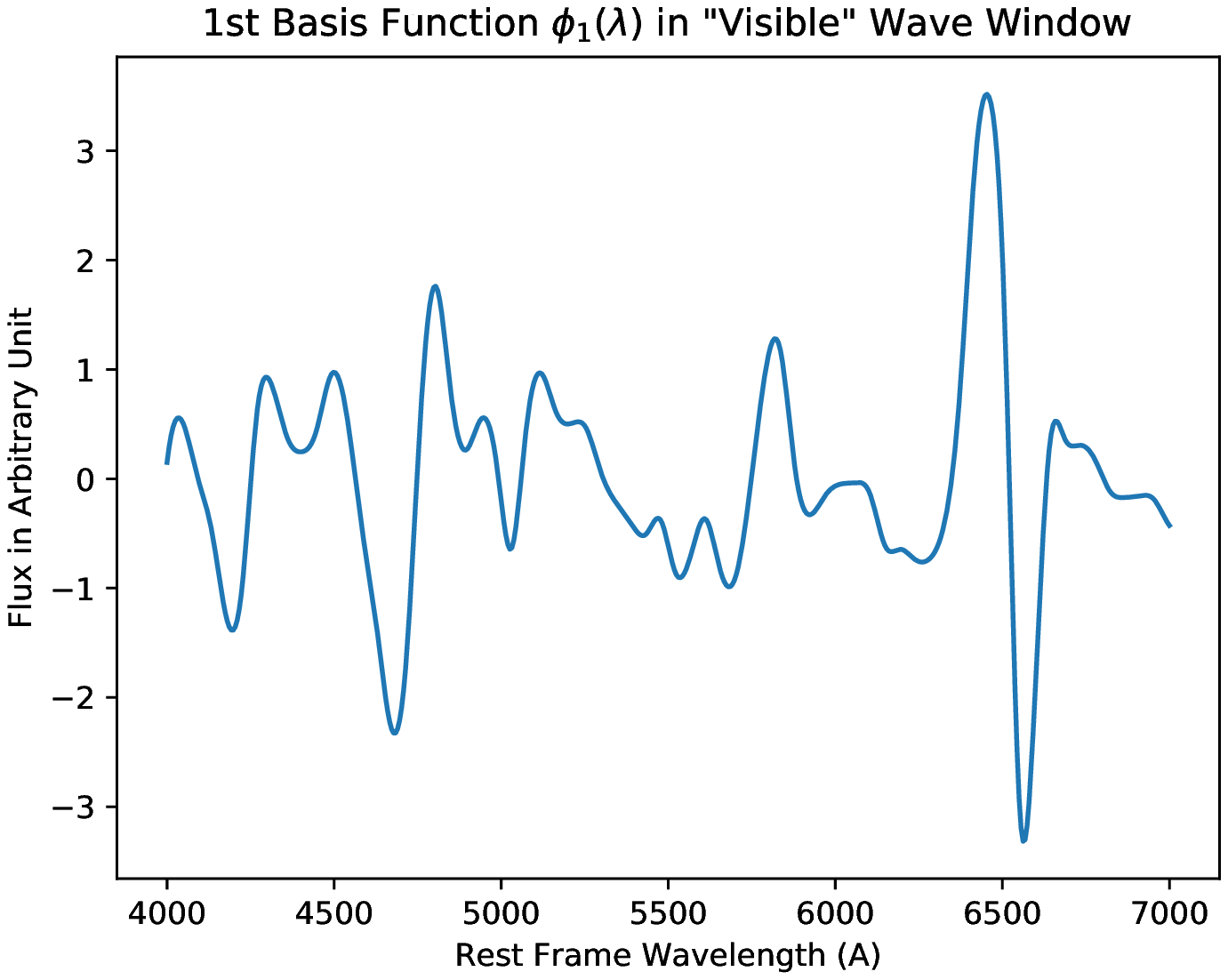}
\endminipage\hfill
\minipage{0.32\textwidth}
  \includegraphics[width=\linewidth]{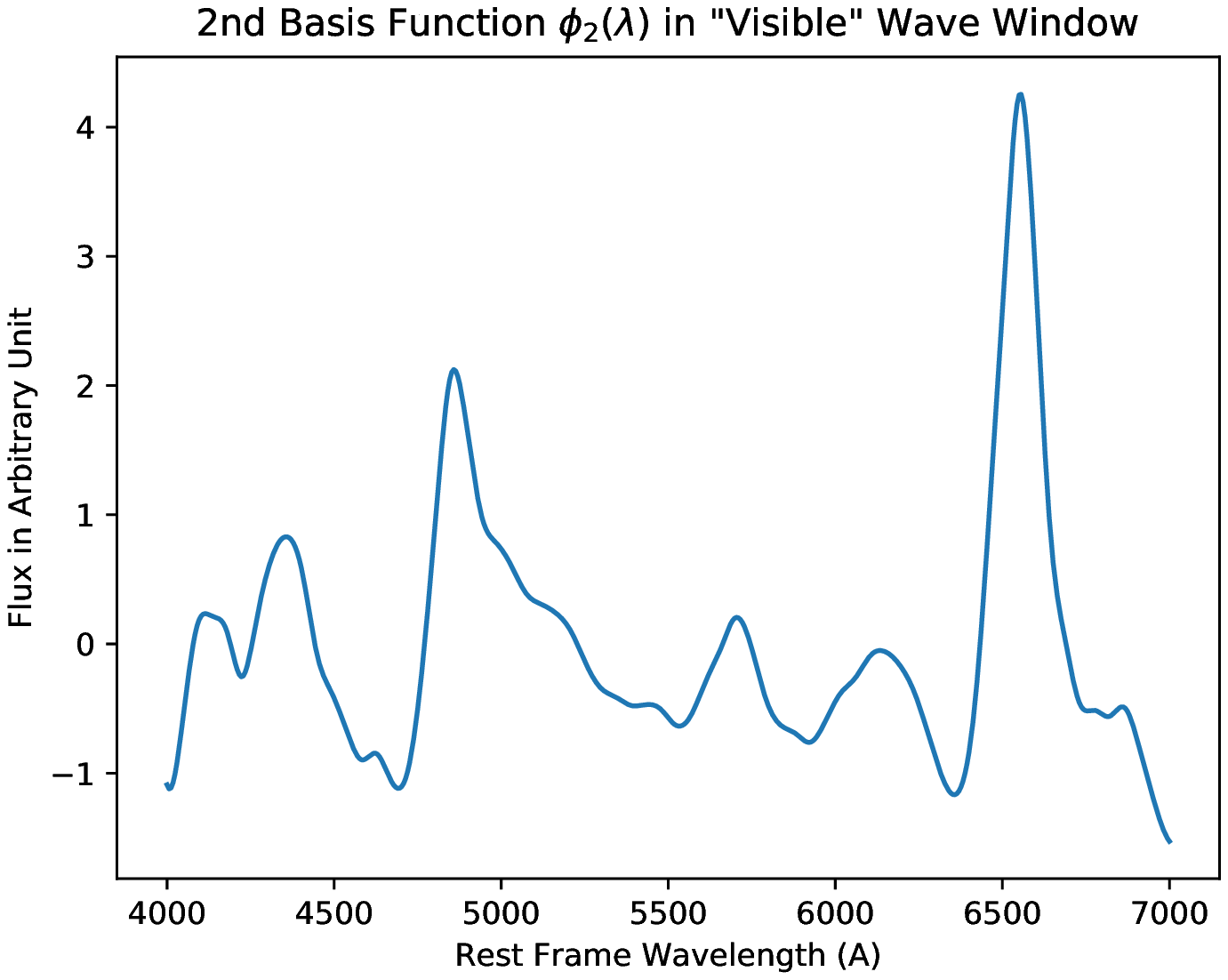}
\endminipage

\minipage{0.32\textwidth}
  \includegraphics[width=\linewidth]{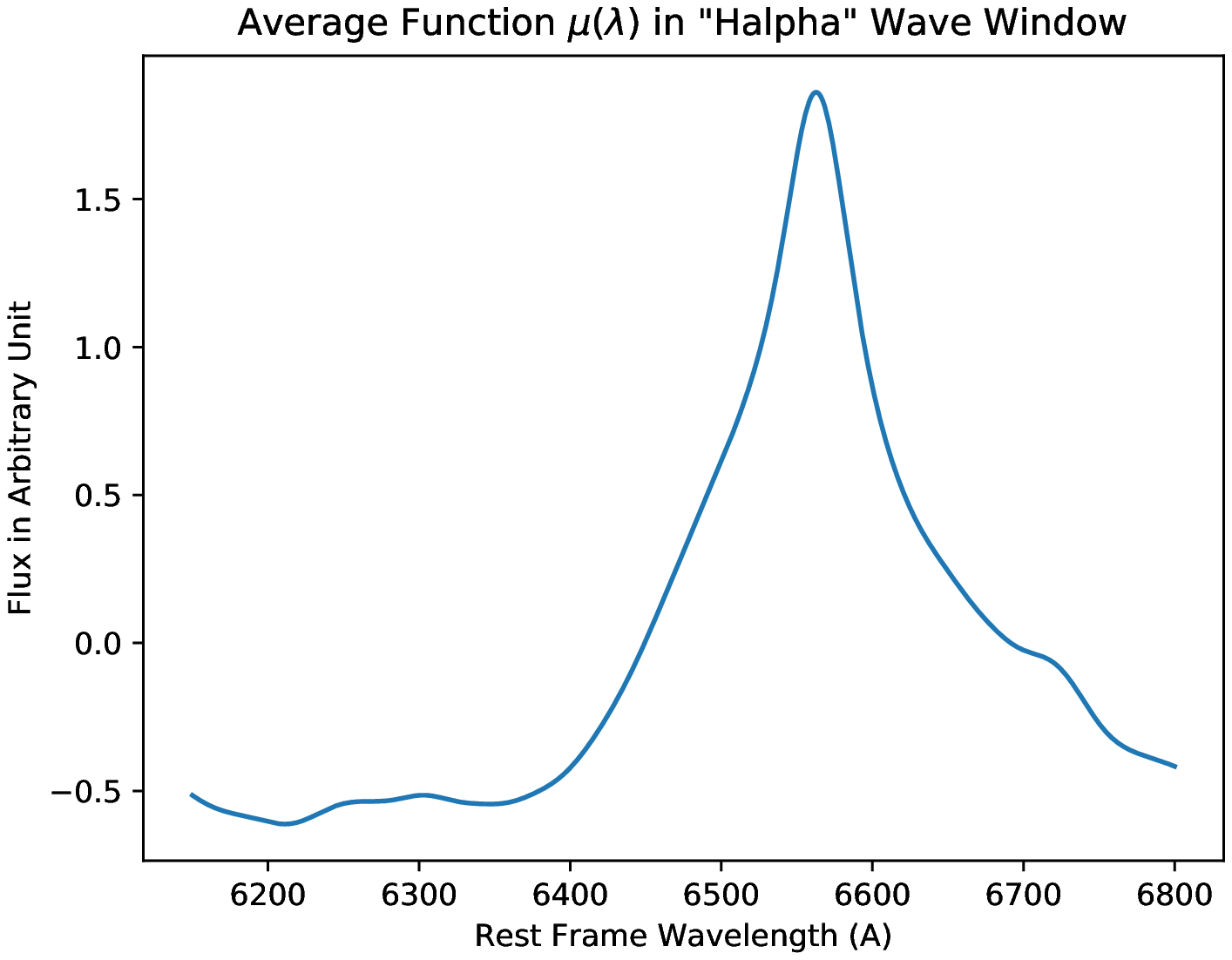}
\endminipage\hfill
\minipage{0.32\textwidth}
  \includegraphics[width=\linewidth]{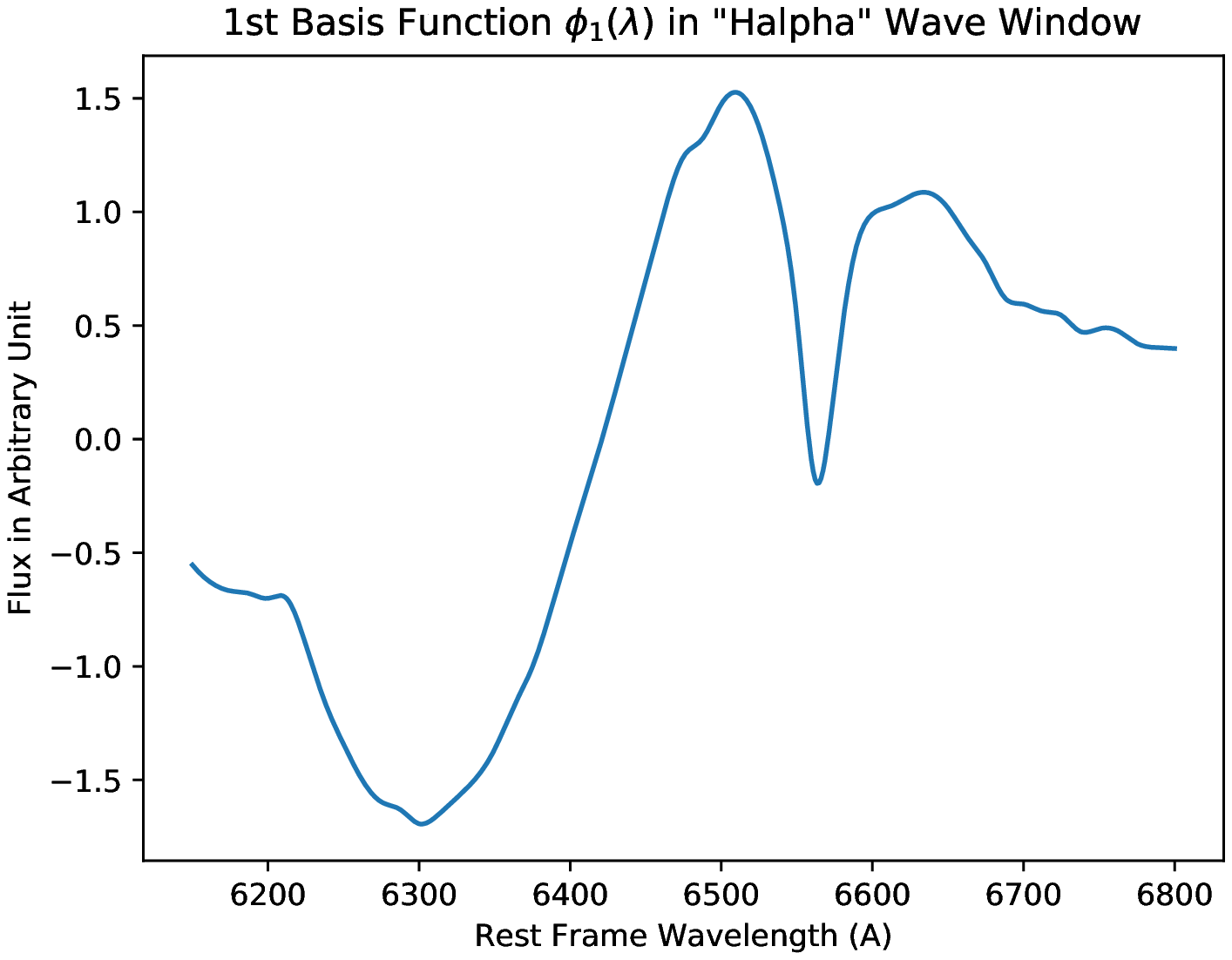}
\endminipage\hfill
\minipage{0.32\textwidth}
  \includegraphics[width=\linewidth]{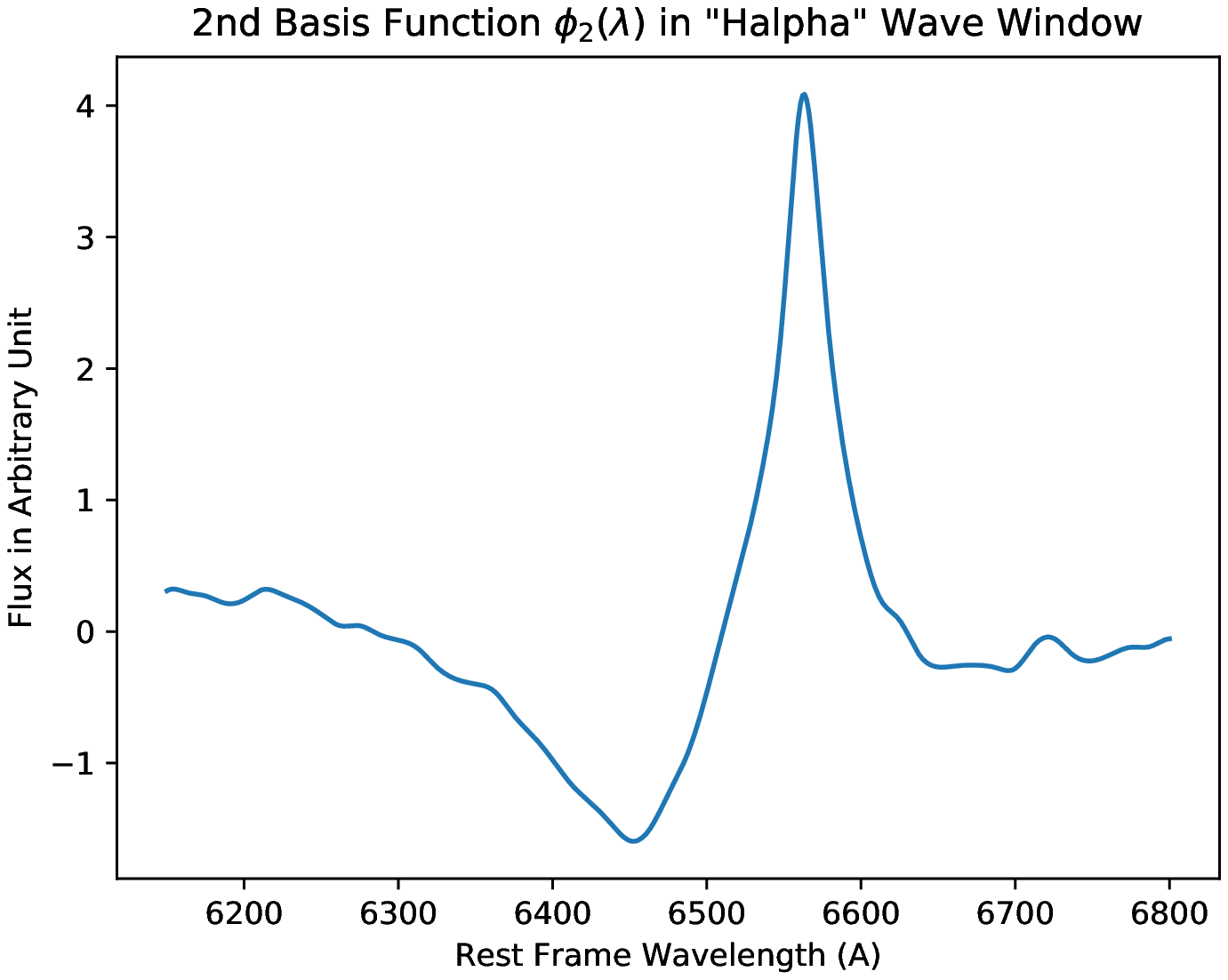}
\endminipage

\caption{Average Function, 1st and 2nd basis function in `Visible' and `$H\alpha$' wave-windows.}\label{fig:fpcs}
\end{figure}

We solved 30 basis functions for each small wave-windows Table~\ref{tab:wavewindow}. 
As for the large wave-windows, 50 basis functions are solved in `Visible' wave-window and 40 basis functions are solved in `Expand' wave-window. We select some basis functions and average functions shown in Fig.\ref{fig:fpcs}. 
Also, we maneged to re-construct some spectra for the comparison with the original data, which is shown in Section~\ref{sec:recon}

\subsection{Support Vector Machine}

From the as-discussed FPCA algorithm, all the spectra can be represented with a series of FPCA scores. 
Althoght some tendencies of different type SNe can be observed in some selected dimensions of FPCA scores' parametric space Section~\ref{sec:twodimplot}, we firstly choose Support Vector Machine (SVM)\citep{svm1,svm2,svm3} for the classification of Type IIL/IIP SNe. 

\begin{figure}
\plotone{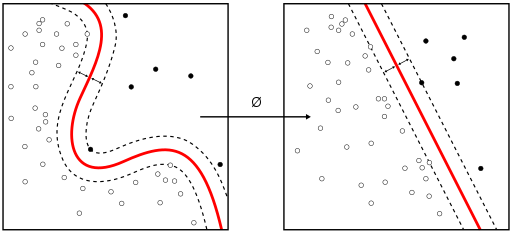}
\caption{An illustrative picture of SVM, By Alisneaky, svg version by User:Zirguezi \url{https://commons.wikimedia.org/w/index.php?curid=47868867}}\label{fig:svm}
\end{figure}

SVM is a popular machine learning method for classification which aims to find the optimal hyperplane to seperate two sets of dots with largest margin distance, as is shown in Fig.\ref{fig:svm}. 
This method is firstly proposed in 1992\citep{svmfirst}

Consider two kinds of nodes in a hyperspace, the target of SVM is to find a hyperplane which could seperate the data, which is written in the form:

\begin{equation}
0=X\cdot \omega + b
\end{equation}

Where X is the coordination on the hyperplane, $\omega$ is the weight of each dimension and b is the biase. 
The dots which are closest to the hyperplane are denoted as support vectors, the distance between the support vector and the hyperplane is: 

\begin{equation}
Min_{i}\left(\frac{\omega\cdot X_i +b}{||\omega||}\right)
\end{equation}

Where $X_i$ is the coordination of the nodes. 
The minimal is reached if the $X_i$ is the support vector. 
To construct an optimize hyperplane, the distances between the support vectors and the hyperplane should be as large as possible, the optimization target: 

\begin{equation}
Max_{\omega,b}\left[Min_{i}\left( \frac{\omega\cdot X_i+b}{||\omega||} \right)\right]
\end{equation}

If the distance between the support vector and the hyperplane is denoted to 1, then the optimization target is:

\begin{equation}
Max(\frac{1}{||\omega||}),\ subject\ to: y_i(\omega\cdot X_i+b>1)
\end{equation}

Where $y_i=\pm1$ is the sign of the nodes, which marks their tags (IIP or IIL). 
Considering the intrinsically not seperable nodes, an slack variable is introduced to gauge the misclassification: 

\begin{equation}
1-\xi_i=y_i(\omega\cdot X_i+b),\ \xi_i>=0
\end{equation}

The cost function for the optimization is written as below: 

\begin{equation}
L(\omega,b,\xi)=\frac{1}{2}||\omega||^2+C\sum_i\xi_i
\end{equation}

In this paper, we use {\tt\string python.sklearn.svm} for SVM classification, radial basis function (RBF) is used for non-linear classification, the kernel function and the distance are written as: 
\begin{equation}
K(x,x')=exp\left(-\gamma\frac{||x-x'||^2}{\sigma^2}\right)
\end{equation}
\begin{equation}
K(\omega,X)+b=exp\left(-\gamma\frac{||x-x'||^2}{\sigma^2}\right)+b
\end{equation}
If no further explanation are given, we set the default parameters $C=3020$ and $\gamma$ automatically generated by the machine, which promises the best performance of SVM algorithm in most classifications in this literature. 

\subsection{Artificial Neural Network}

A simple Artificial Neural Network (ANN) is composed of three parts: input layer, hidden layers and the output layer\citep{deeplearning}. 
Nodes in adjacent layers are linked with different weights. 
In the ANN, data are propogated from the imput layer to the output layer with the equation:

\begin{equation}
z_i=\omega_i\cdot y_i+b
y_{i+1}=f_{i+1}(z_i)
\end{equation}

Where $\omega_i$ is the weight, b is the biase and $f_{i+1}(z_i)$ is the activition function of nodes in each layer. 
The output layer will give the probablistic classification upon a certain input data. 
In this paper, we use recitfied linear unit ($f(z)=max(0,z)$) as the hidden layer's activition function, and sigmoid function ($f(z)=1/(1+exp(-x))$) as the output layer's activition function. 

The weight and bias parameters are trained using backpropogation procedure. 
In this paper, we choose binary cross entropy\citep{crossentropy} as the loss function, which is written as: 

\begin{equation}
L(y_r,y_p)=-\sum_i \left[y_r\ln(y_p)+(1-y_r)\ln(1-y_p)\right]
\end{equation}

Where $y_r$ is the true probablity 0 or 1, $y_p$ is the probablity given by the neural network. 
For each training epoch, the upload of weight and biase are uploaded upon the gradient of the loss function, $g_\omega=\frac{\partial L}{\partial \omega},\ g_b=\frac{\partial L}{\partial b}$
In this paper, we use Adam algorithm\citep{adam} for optimization, from which the weights and biase are updated with:

\begin{eqnarray}
m_t=\beta_1m_{t-1}+(1-\beta_1)g\\
v_t=\beta_2v_{t-1}+(1-\beta_2)g\\
\hat{m_t}=m_t/(1+\beta_1^t)\\
\hat{v_t}=v_t/(1+\beta_2^t)\\
\theta_{new}=\theta_{old}-\alpha \hat{m_t}/(\sqrt{\hat{v_t}}+\epsilon)
\end{eqnarray}

Where $\theta$ is weight or biase, $m_t$ and $v_t$ are the t-th epoch first and second momentum, $\epsilon=10^{-8}$ is set to avoid overfull. 
We use the default decay parameters $\beta_1=0.9,\ \beta_2=0.999$ and the learning rate $\alpha=0.001$.  

In this paper, we use the integrated deep learning package {\tt\string python.keras} for ANN training and testing. 
Regulizations are not introduced considering they cannod increase the performance in this paper.

\begin{figure}[ht!]
\includegraphics[width=400pt, height=200pt]{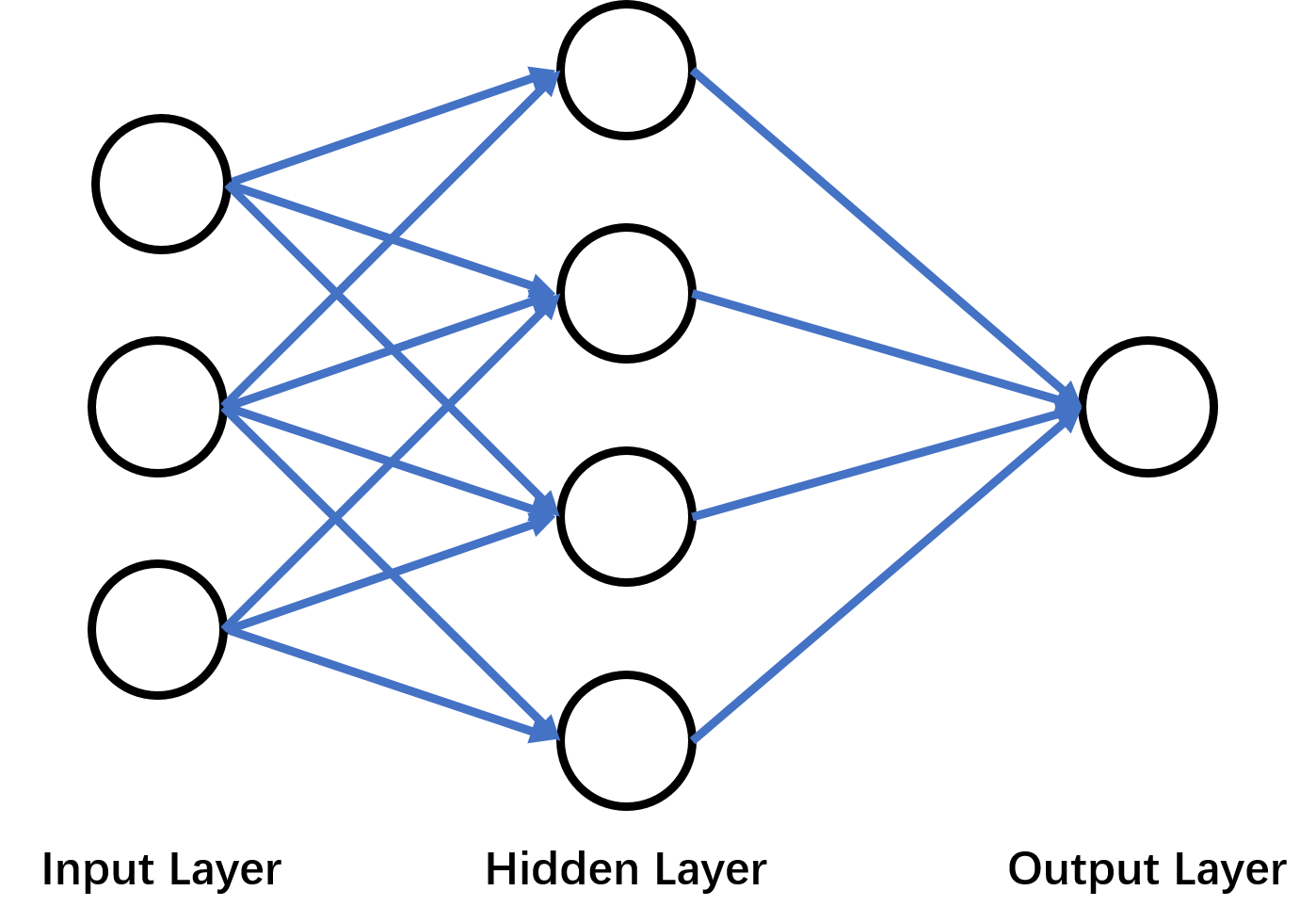}
\caption{The Formation of Artificial Neural Network. Here, only one neurone is used in output layer for dichotomy classificatin of Type IIP/IIL SNe.}\label{fig:neurlstructrue}
\end{figure}

\subsection{Performance evaluation}\label{sec:performance}

In each trainning and testing process, we randomly discard some Type IIP SNe at the outset, in order to keep a same number of Type IIP/IIL spectra in our dataset. 
Secondly, the dataset is splited into training set and testing set, which contains 80\% and 20\% of the data. 
When the classifier has trained on the training set, we use F1-Score to evaluate the performances of our classifiers. the definition of F1-Score is as follows\citep{f1score}: 

\begin{equation}
{\rm Precision=\frac{True Positive}{True Positive+False Positive}}
\end{equation}
\begin{equation}
{\rm Recall=\frac{True Positive}{True Positive+False Negative}}
\end{equation}
\begin{equation}
{\rm F1-Score=\frac{2}{\frac{1}{Precision}+\frac{1}{Recall}}}
\end{equation}

Where true positive is the number of `positive' spectra retrived by the classifier, true negative is the number of `negative' spectra that didn't retrived by the classifier, false positive is the number of `negative' wrongly retrived by the classivier and false negative is the number of `positive' spectra didn't retrived by the classifier. 
We calculated the precision, recall and F1-Score of Type IIP and Type IIL seperately, then adopt the average of two types. 

Because the number of Type IIL/IIP spectra are too small and biased, as is shown in Table~\ref{tab:wavewindow}, we adopt a cross validation method to evaluate our classifiers' performance. The training- and testing-process will repeat several times and the dataset are re-splitted before each training process. 
Finally, we calcuate the average precision, recall and F1-Score to evaluate the classifier. 
In addition, the standard error of precision, recall and F1-Score are also calculated in order to evaluate the stability of classifiers.

\section{results and discussion}\label{sec:result}

\subsection{The Performance of Classifiers in One Wave-Window}\label{sec:onewavewindow}

We firstly test the performance of SVM using only one wave-window and calculate the F1-Score, the results are shown in Table~\ref{tab:svmone}. 
In this trial, only first 30 FPCA scores of the `Visible' and `Expand' are selected, to perform an unbiased test to compare the performance of smaller and larger wave-windows.  
We notice that both `Visible' and `Expand' wavw-windows have a relatively high F1-Score, which reaches to 0.815 and 0.83. 
In contrast, the `NaMg' wave-window's F1-Score is the lowest among others. 
Moreover, we notice the F1-Score of `$H_\alpha$' window reaches 0.809, which is the third highest among 11.

Additionally, we utilized several ANN models for the classification using 30 FPCA scores in one wave-window, the results of which is shown in Table~\ref{tab:neuro}. 
A simple three-layered neural network is constructed in {\tt\string python.keras} framework, the parameters are listed in Table~\ref{tab:neuropara1}. 
In the list of classification results Table~\ref{tab:neuro}, only the scores of best-performed model are shown in the table. 
The F1-Score of ANN is higher than SVM's in all wave-windows. 
Notably, the F1-Score of `$H_\alpha$' wave-window, which reaches 0.849, is the highest among other 8 small wave-windows. 

Although large wave-windows are not excel at re-constructing every details of a spectrum Section~\ref{sec:methods}, its performance is higher than small wave-windows, we suggest this phenomenon is caused by the normalization process in pro-processing the spectra. 
For the precision in solving FPCA basis functions, every wave-windows are normalized seperately. 
As a consequense, the flux ratio between each small wave-windows are omitted. 
Considering this part of information is directly linked to the element ratios in SNe, it is persumable to obtain a worse result when solely small wave-windows are utilized. 
In contrast, a comprehensive wave-window contains multiple spectral lines, and the information about the element ratios are preserved, which promises a better performance.

\begin{deluxetable}{c|c}[ht!]
\tablecaption{ANN parameters for one-window classification}\label{tab:neuropara1}
\tablehead{\colhead{Items}&\colhead{Keys}}
\startdata
Input Layer & 30\\
Hidden Layer & See the Table~\ref{tab:neuro}\\
Activition Function & ReLU\\
Output Layer & 1\\
Activition Function & sigmoid\\
Optimisztion Method & AdamBoost\\
Regularization & None\\
\enddata
\end{deluxetable}

\begin{figure}
\plotone{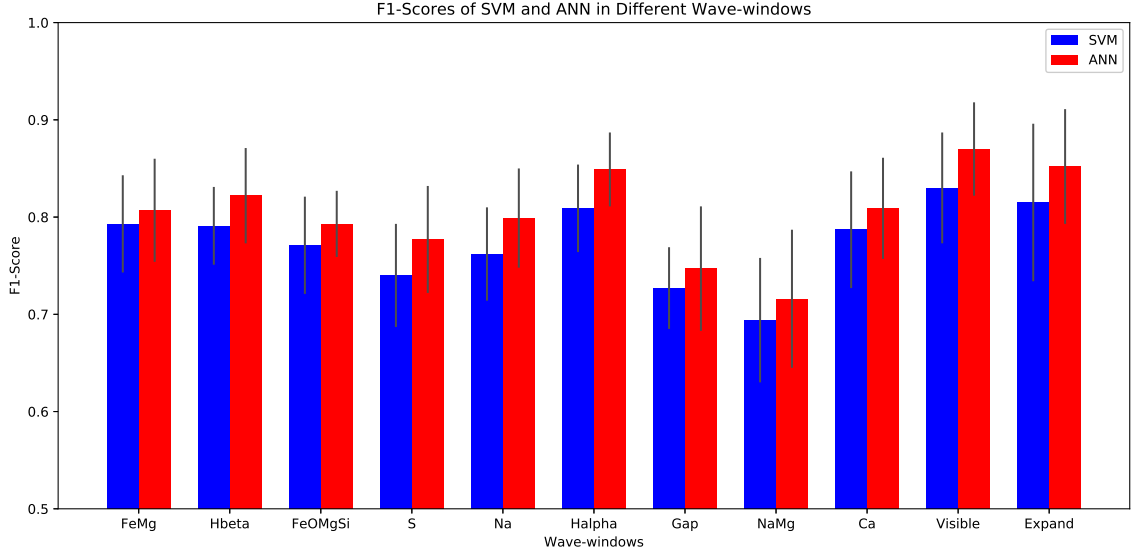}
\caption{The F1-Score in SVM and ANN of every wave-windows. Error bars are plotted in the picture. The F1-Score of ANN are generally higher than SVM's. Data from Table~\ref{tab:neuro}, Table~\ref{tab:svmone}}
\end{figure}

\subsection{Performance with More Basis Functions}\label{sec:performancewithless}

\begin{figure}[ht!]
\plottwo{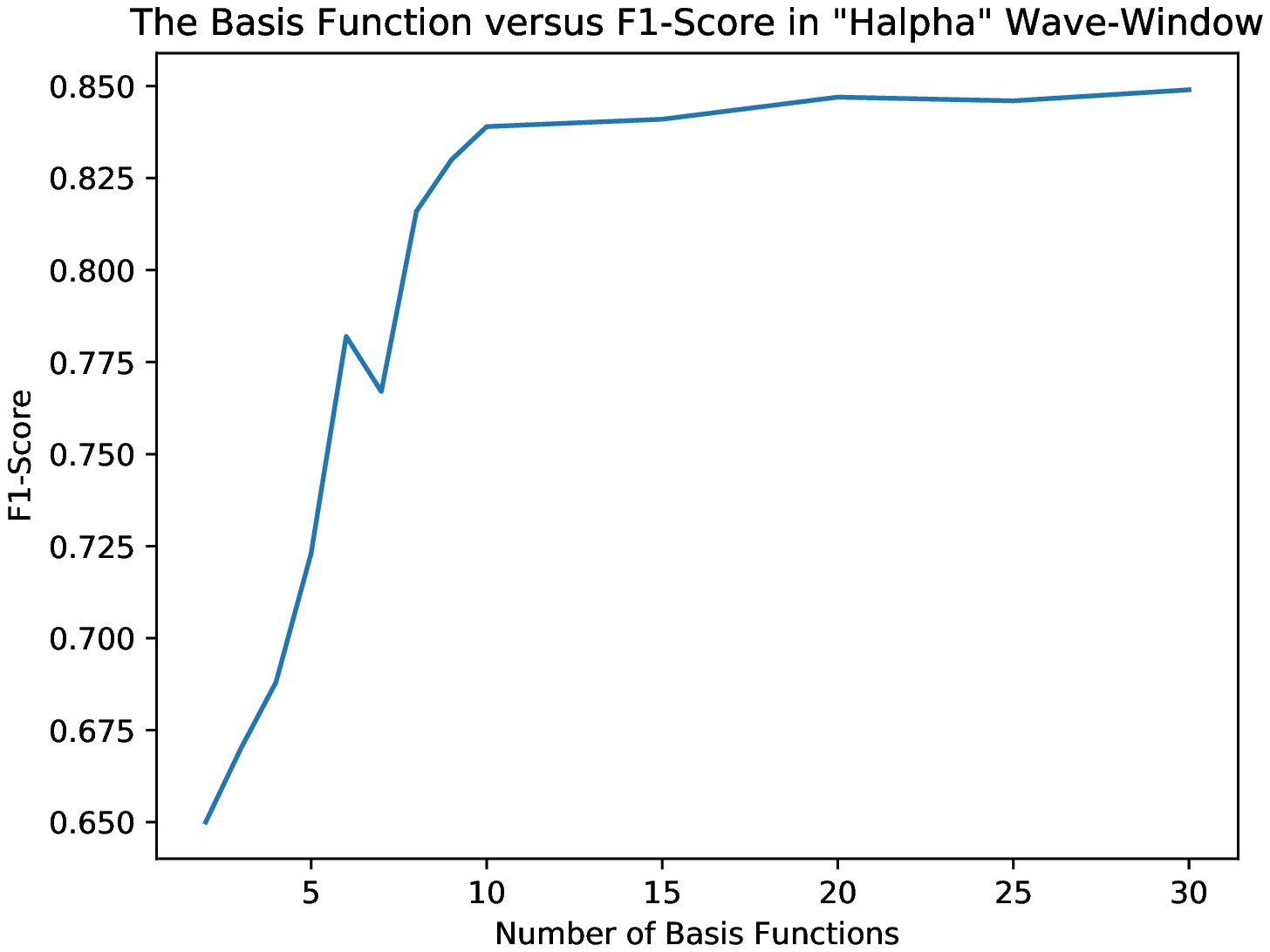}{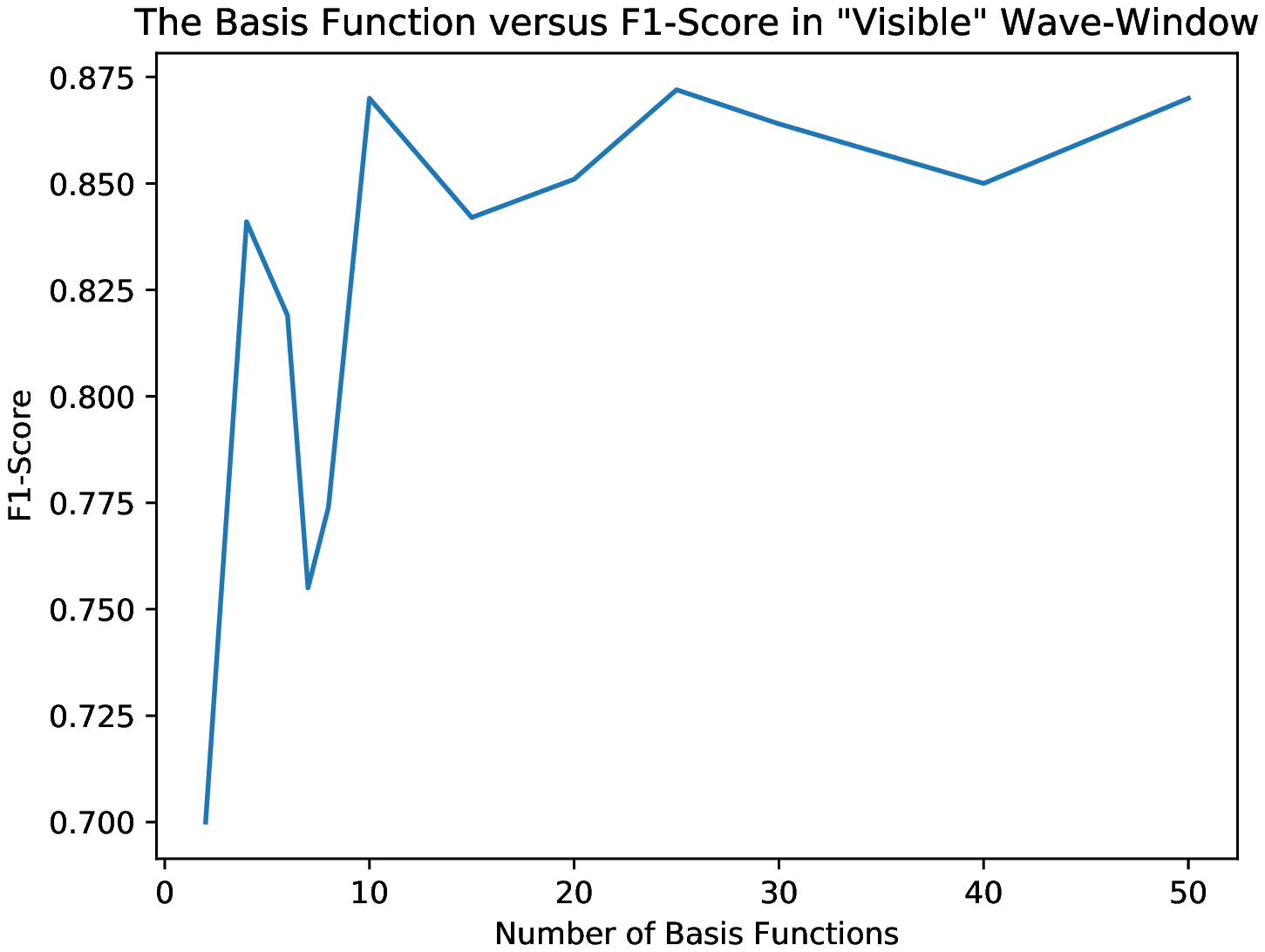}
\caption{The relation between the FPCA scores dimension and the F1-Score in `$H_\alpha$' and `Visible' wave-windows. }\label{fig:changedimension}
\end{figure}

With more basis functions, it is possible to re-construct the spectra with a higher fidelity, albeit more computational costs in solving basis functions. 
However, whether more FPCA scores will increase the accuracy in classification remains a question. 
In this section, we tried to change the number of basis functions in FPCA analysis, and discuss the minimal number of basis functions for the classification. 
Relating to the F1-Scores in Section~\ref{sec:onewavewindow}, we choose the best small wave-window and the best large wave-window, `$H_\alpha$' and `Visible' for the following discussion. 
The models of ANN are the same in Section~\ref{sec:onewavewindow}
We plotted our results into Fig.\ref{fig:changedimension}, the relating data are shown in Section~\ref{sec:data}. 
In `$H_\alpha$' wave-window, F1-Score reaches 0.839 when 10 basis functions are utilized, while the F1-Score in `Visible' wave-window for 10 basis functions is 0.842. 
If the dimension is larger than 10, the F1-Scores of two wave-windows are more than 90\% of the wave-windows' best F1-Score, while the outcome slopes when the dimension is smaller than 10. 
Moreover, we can observe fluctuations in `Visible' bands, indicating the instability of the ANN model in the chosen wave-window, which is also confirmed by the relatively large variation ($0.5-0.9$) of F1-Score. 
Accordingly, we suggest 10 basis functions is adequete for the classification of Type IIP/IIL SNe. 

Moreover, we availed the most data and attempt to test the stratosphere. 
We tested the classifiers' performance in using 9 small wave-windows or all 11 wave-windows, the total number of FPCA scores of which reaches 270 and 360. 
So, the imput dimension is 270 or 360. 
Using the default parameters in SVM $C=3020$ still performs well. 
However, we found that using neural-network with 90 or 30 nodes in the hidden layer performs the best. 
The results are listed below. 

\begin{deluxetable}{ccc|ccc|ccc}
\tablecaption{The Best-Performed Classifiers}
\tablehead{
\colhead{Classifier} & \colhead{Number of Wave-Windows} & \colhead{Hidden-Layer Nodes} & \colhead{Precision} & \colhead{Recall} & \colhead{F1-Score} 
& \colhead{$\sigma_P$} & \colhead{$\sigma_R$} & \colhead{$\sigma_F$}
}
\startdata
SVM & 9 & / & 0.829 & 0.831 & 0.824 & 0.071 & 0.071 & 0.074\\
SVM & 11& / & 0.831 & 0.829 & 0.824 & 0.078 & 0.079 & 0.080\\
ANN & 9 & 90 & 0.856 & 0.858 & 0.852 & 0.053 & 0.054 & 0.056\\
ANN & 11&30 & 0.886 & 0.886 & 0.881 & 0.054 & 0.059 & 0.057\\
\enddata
\tablecomments{In this scenario, the parameter for SVM C=3020. }
\end{deluxetable} 

Comparing to the SVM in one wave-window Table~\ref{tab:svmone}, the results in using multiple wave-windows did not increase whatsoever, probably because the higher dimension of features infects SVM's performance \citep{highdimsvm1,highdimsvm2}. 
In using ANN for classification, we found the F1-Score reaches 0.881, which is the best among other classification methods in this paper. 

\section{summary}\label{sec:summary}

We decomposed Type II SNe spectra into different basis functions via FPCA algorithm. 
Based on the FPCA scores of every spectra, we have trained SVM and ANN to classify Type IIP and IIL SNe, two types of which has scant spectroscopic discrepancies. 
The spectra are divided into 9 small-size wave-windows and 2 large-size wave-windows for FPCA analysis respectively. The best F1-Score we got is 0.881. 
It is, for the first time, the effectively spectroscopic classification of Type IIP/IIL SNe. 

Using the FPCA algorithm in different wave-windows, we successfully compressed the data of some certain spectral lines into 30-50 FPCA scores with minimal data loss. 
This processing trick do not only preserves the most informations about the elements explosion speed, but also enables us to apply some simple machine-learning models for classifications. 

Moreover, we notice that only using a small-scale wave window at 6150-6800 \AA, which covers the P Cygni profile of $H_\alpha$ line, outperforms to other spectral features in classification with its F1-Score up to 0.849, probable because of the higher signal-to-noise ratio. 
We suggest FPCA analysis directly on spectral line of a specific element could be a new approach in SNe classification analysis. 

In the future, we will investigate the relationship between the basis functions (the principal components in spectra) and the explosion profile of SNe to expand the classification method on Type I SNe.
Moreover, we will integrate other informations, such as temperature and the color evolution of SNe into the model for a higher accuracy. 

\acknowledgements

Xingzhuo Chen thanks to Prof. Avishay Gal-Yam (Weizmann Institute of Science) and Prof. Lifan Wang (Purple Mountain Observatory) for supportive discussion. 
We thanks Weizmann Interactive Supernova data REPository (WISeREP) \url{https://wiserep.weizmann.ac.il/} and Transient Name Server (TNS) \url{https://wis-tns.weizmann.ac.il/} for the data. 

\software{python,keras,scikit-learn,R-fpca\citep{fpca3}}

\bibliographystyle{authordate1}
\bibliography{Startup-SN.bib}

\section{methods}\label{sec:methods}

\begin{figure}[ht!]
\plottwo{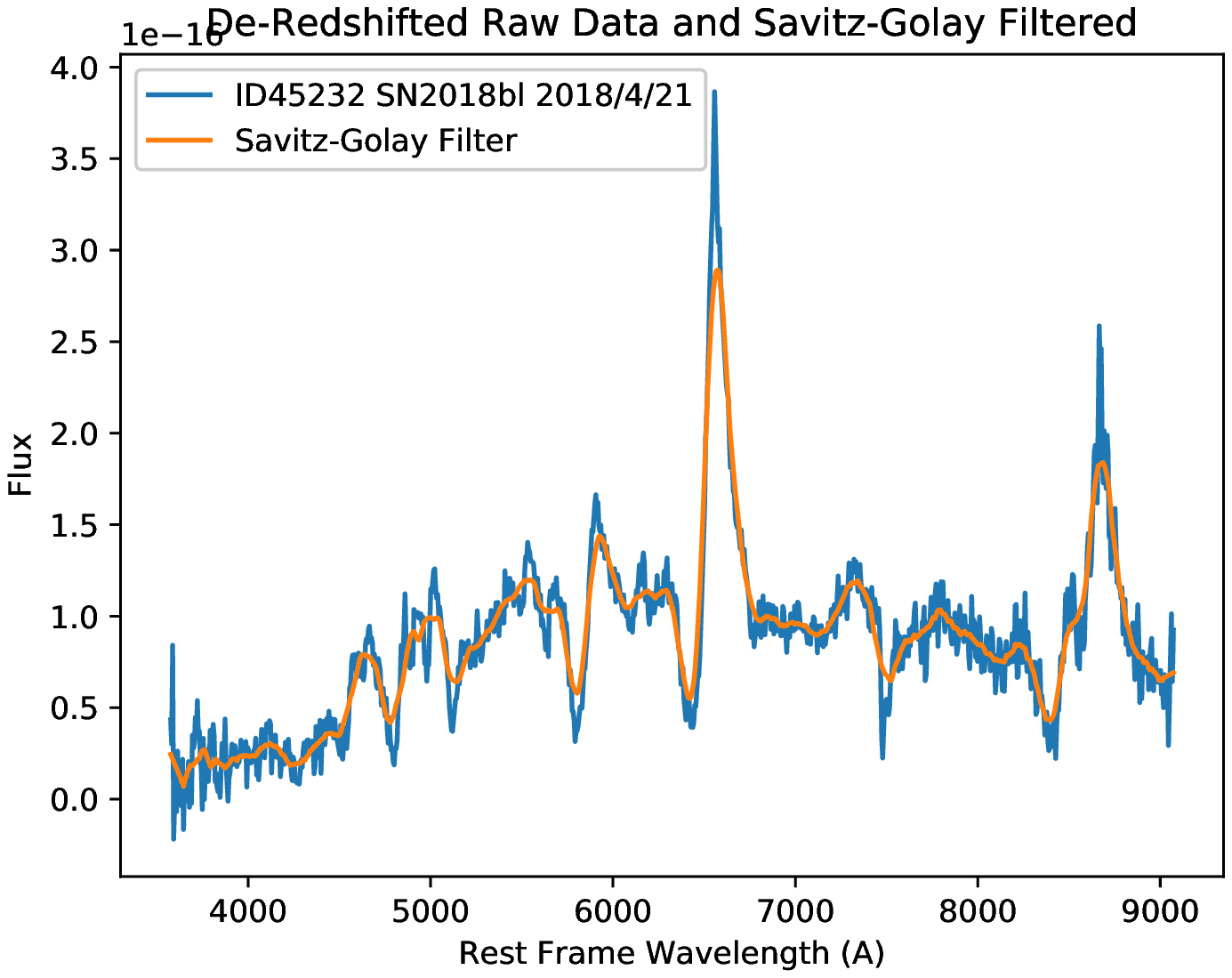}{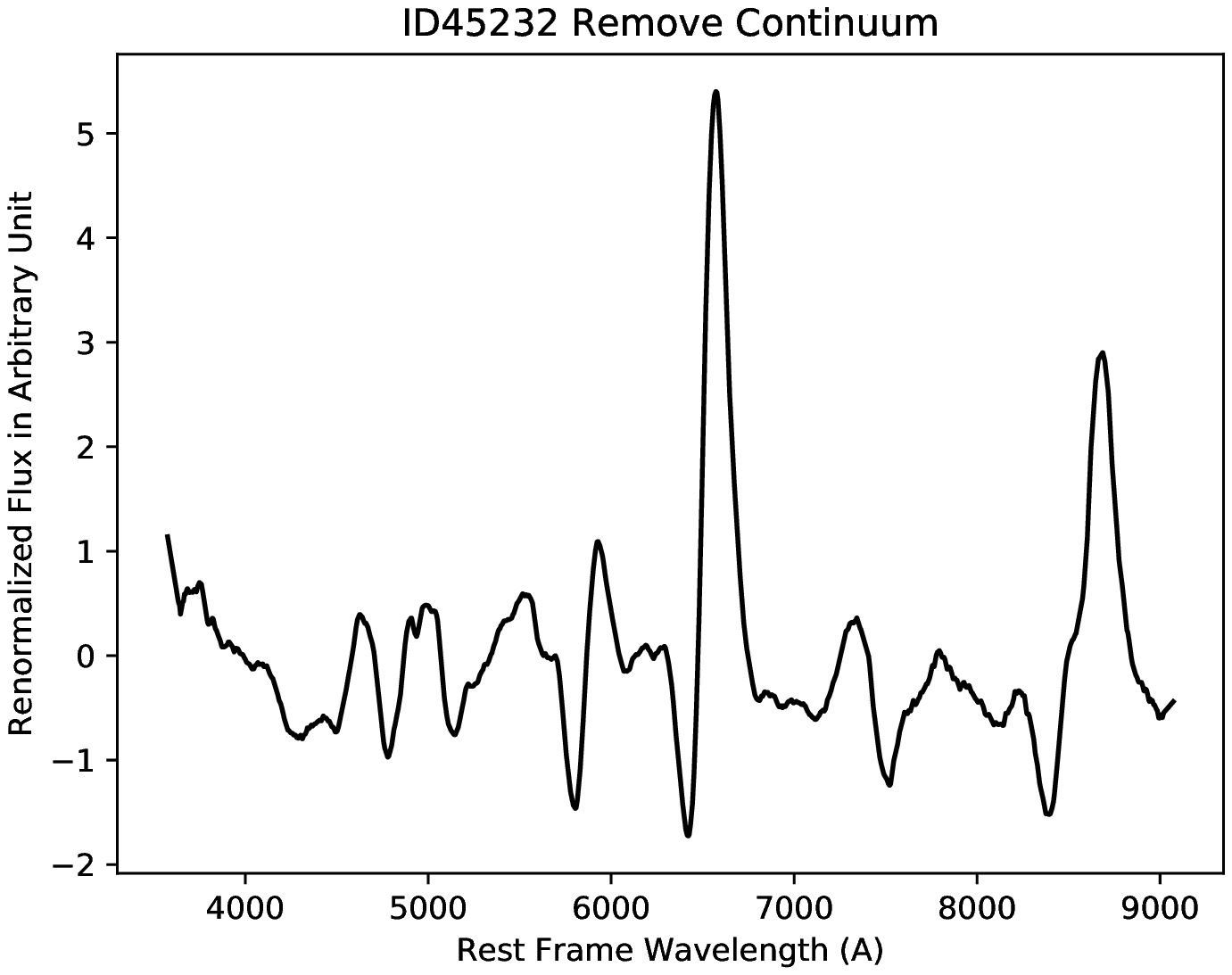}
\caption{An example spectrum before and after the Savitzky-Golay filter. The window and the order of Savitzky-Golay filter in this plot is $2\times int(10/\Delta \lambda)+1$ and 1, $\Delta \lambda$ is the average wave interval of each spectra. ID45232 is the sppectra number in WISeREP.}\label{fig:sg}
\end{figure}

The raw spectrum from telescopes will firstly remove the redshift by using the equation $\lambda_{RF}=\lambda_{obs}/(1+z)$, where $\lambda_{RF}$ is the wavelength in the rest-frame, $\lambda_{obs}$ the wavelength in the observer frame. 
Then, a Savitzky-Golay filter is applied to smooth the spectra, which is shown in Fig.\ref{fig:sg}. 
Considering the black body radiation in the spectra, we adopt an univariate spline method {\tt\string python.scipy.UnivariateSpline} for continuum removal, instead of fitting the blackbody radiation spectra, because the iron-blanketing effect may cause extra continuum absorption at the wavelength small than 5000 \AA \ \citep{ironblanketing}. 
Furthermore, every spectra' fluxes are re-normalized to average equal zero and variance equal 1. 
After this step, we have obtained the spectra preserves most informations of their element compositions. 

In the whole pre-processing process, we omitted the color and the temperature information in the spectra, but most informations about the element composition (the position and the depth of spectral lines) and the explosion speed (the P-Cygni profile and the psudo-Equivalent-Width of spectral line) are preserved. 
Although the classification results are desirable, we will attempt to integrate these information into the model in the future researches. 

\section{FPCA fidelity discussion}\label{sec:recon}

\begin{figure}[!htb]
\minipage{0.32\textwidth}
  \includegraphics[width=\linewidth]{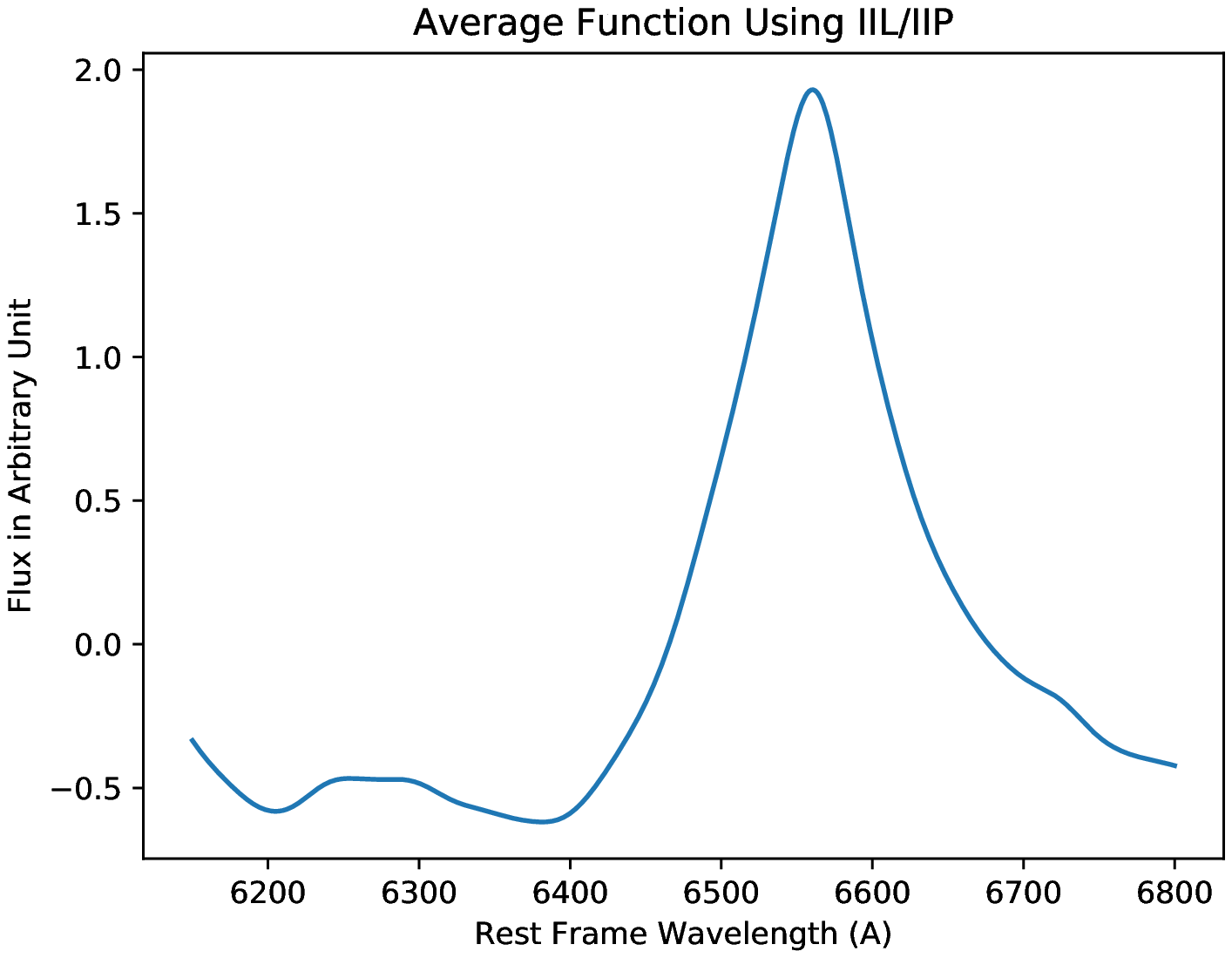}
\label{fig:ex1andex3}
\endminipage\hfill
\minipage{0.32\textwidth}
  \includegraphics[width=\linewidth]{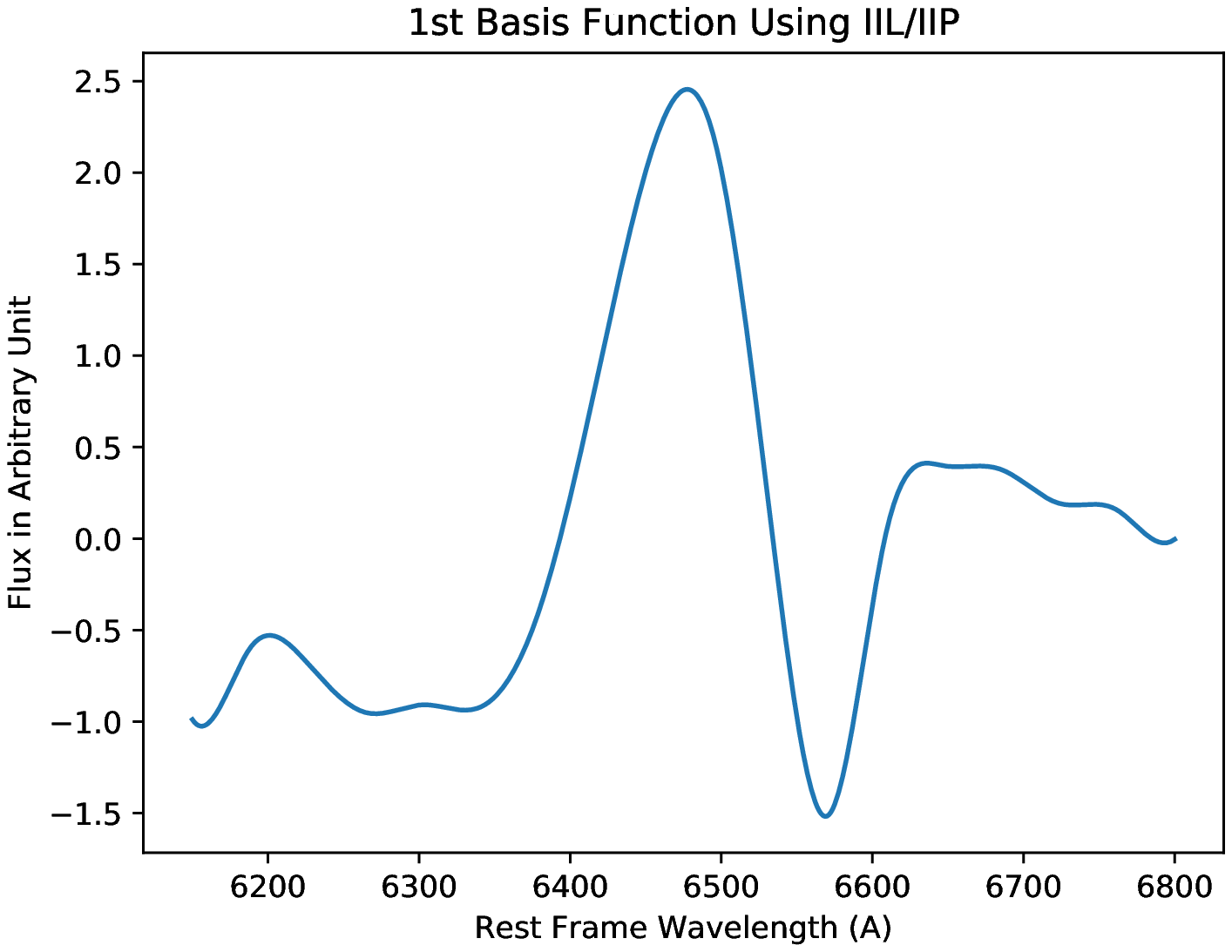}
\label{fig:IIPLh1h3_one}
\endminipage\hfill
\minipage{0.32\textwidth}
  \includegraphics[width=\linewidth]{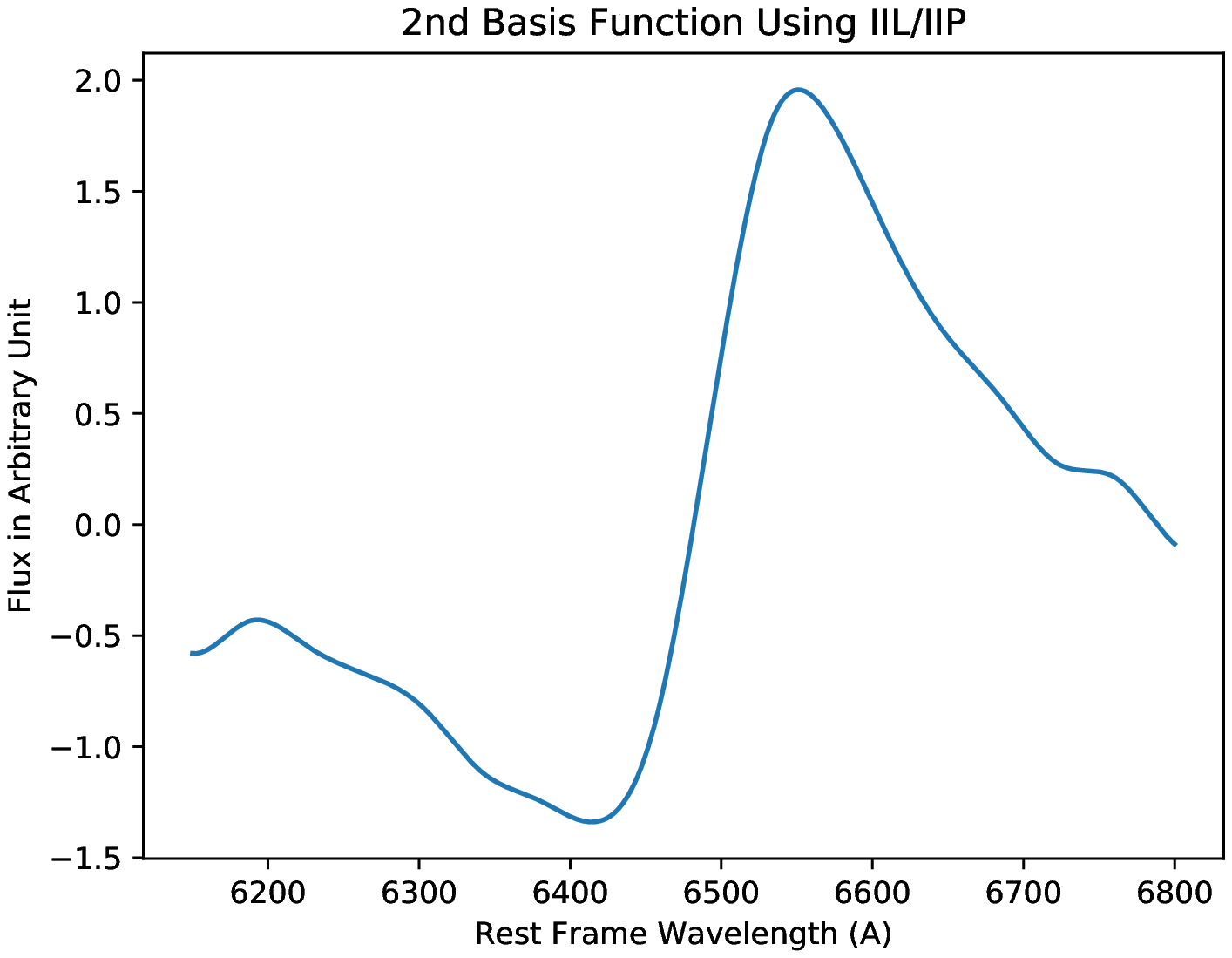}
\label{fig:timeevolution}
\endminipage
\label{fig:IILPbasis}
\end{figure}

Considering our computer compatibility, we didn't use all the spectra we downloaded for FPCA analysis. 
Instead, we randomly choose 15\% of them to solve the basis functions. 
According to some comprehensive research on the Type II SNe\citep{faranIIPL,faranIIPL2} which suggesting the potential continuum between the subtypes of Type II SNe, II/IIb/IIn are also used in FPCA analysis.
With these basis functions, the FPCA score of the rest spectra can be calculated from Eq.\ref{eq:convolution}.
The basis functions of each time may varies, however, we didn't observe basis functions from different spectra affecting the performance of classifiers or the fidelity when re-constructing the spectra. 
We also choose only Type IIP and Type IIL spectra for FPCA analysis, the average function and first two basis functions in `$H_\alpha$' window is shown in Fig.\ref{fig:IILPbasis}. 
Comparing to Fig.\ref{fig:fpcs}, the sharp feature in the first basis function at around 6700 \AA is absent, which is probably contributed to the relatively narrow $H_\alpha$ line in Type IIn SNe' spectra. 
Nonetheless, with an unknown reason, using these sets of basis functions didn't increase the performance in classifing Type IIP/IIL SNe. 

\begin{figure}[ht!]
\plottwo{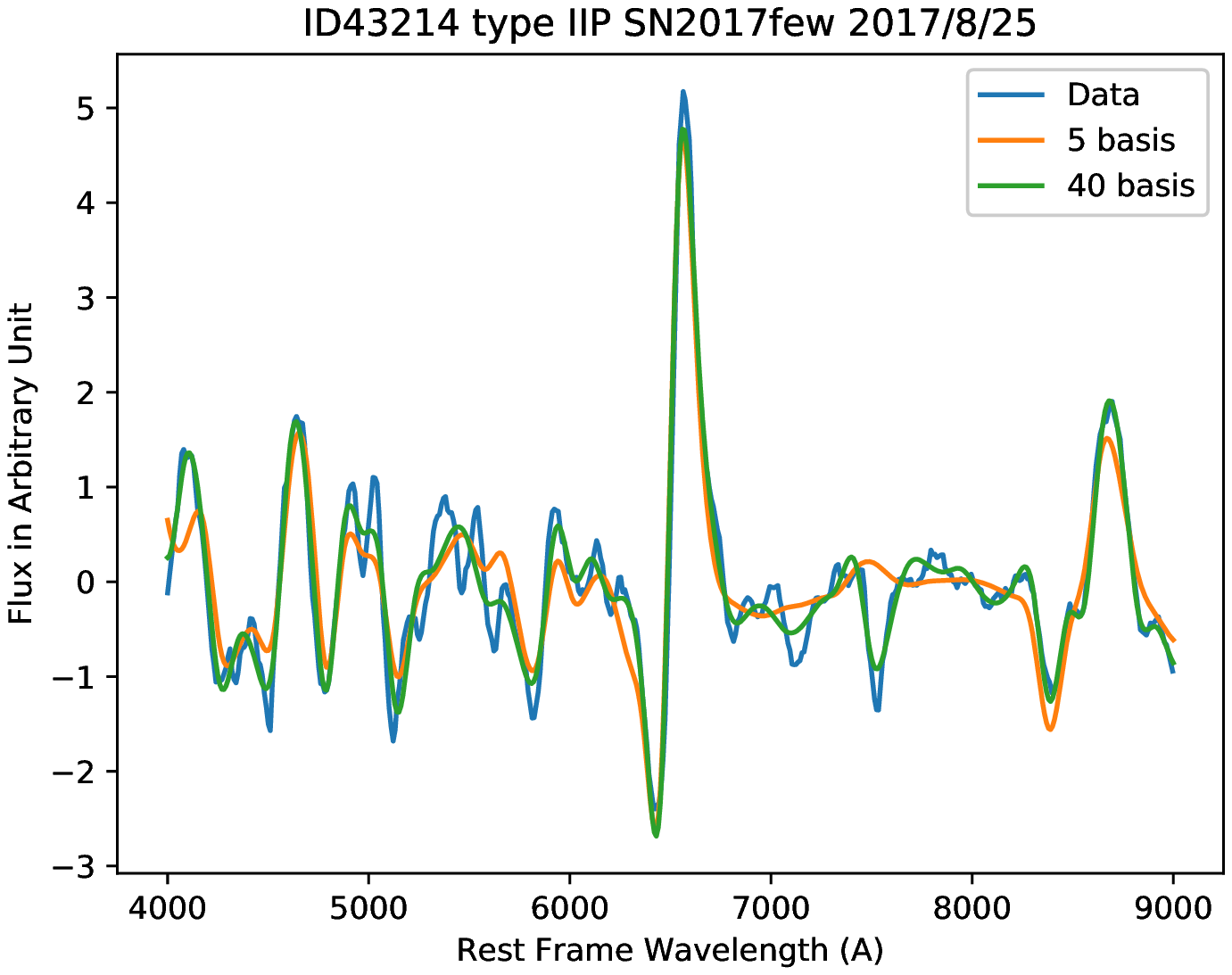}{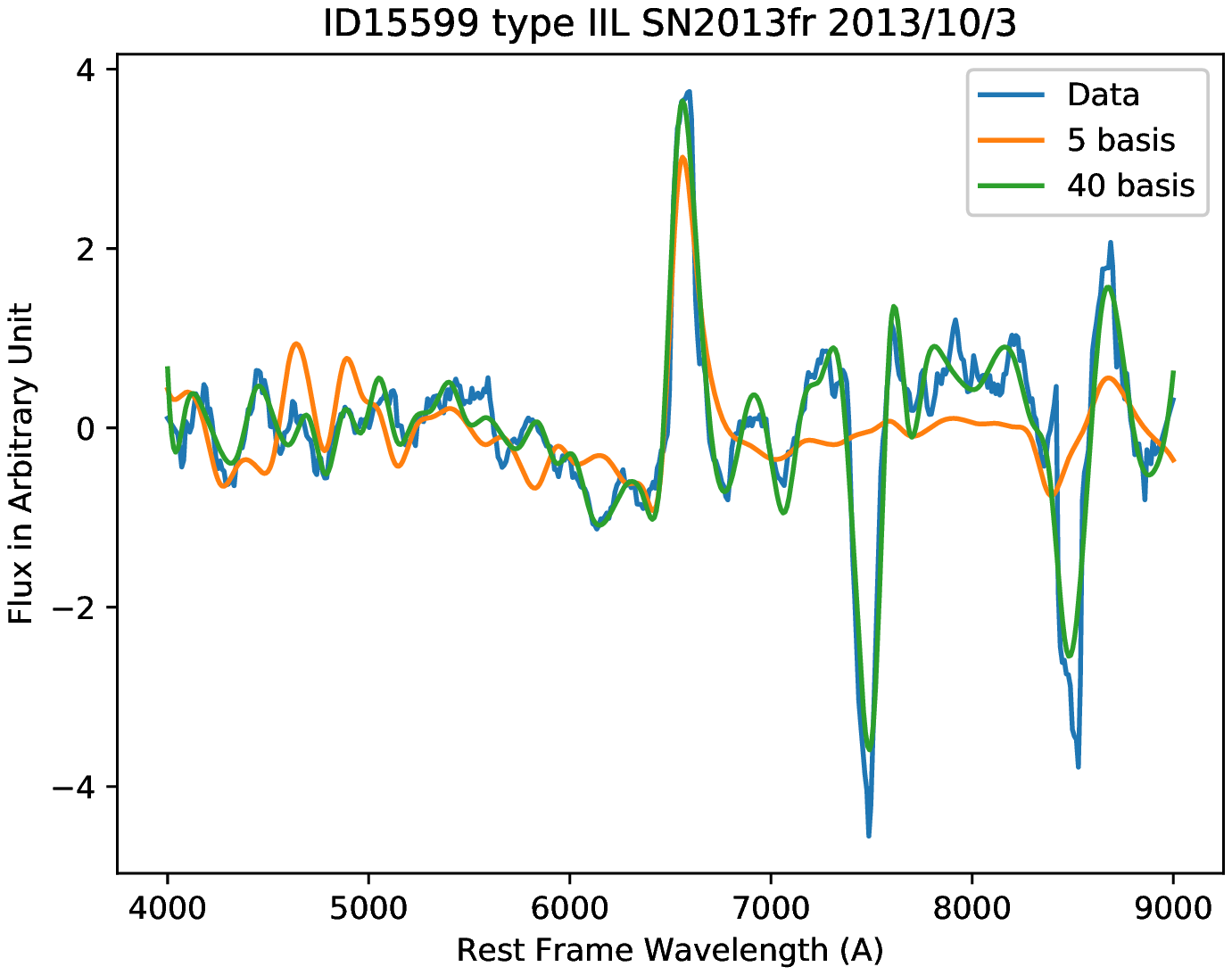}
\caption{The fitting results of two example spectra, ID number is the spectrum number in WISeREP.}\label{fig:fitresult}
\plottwo{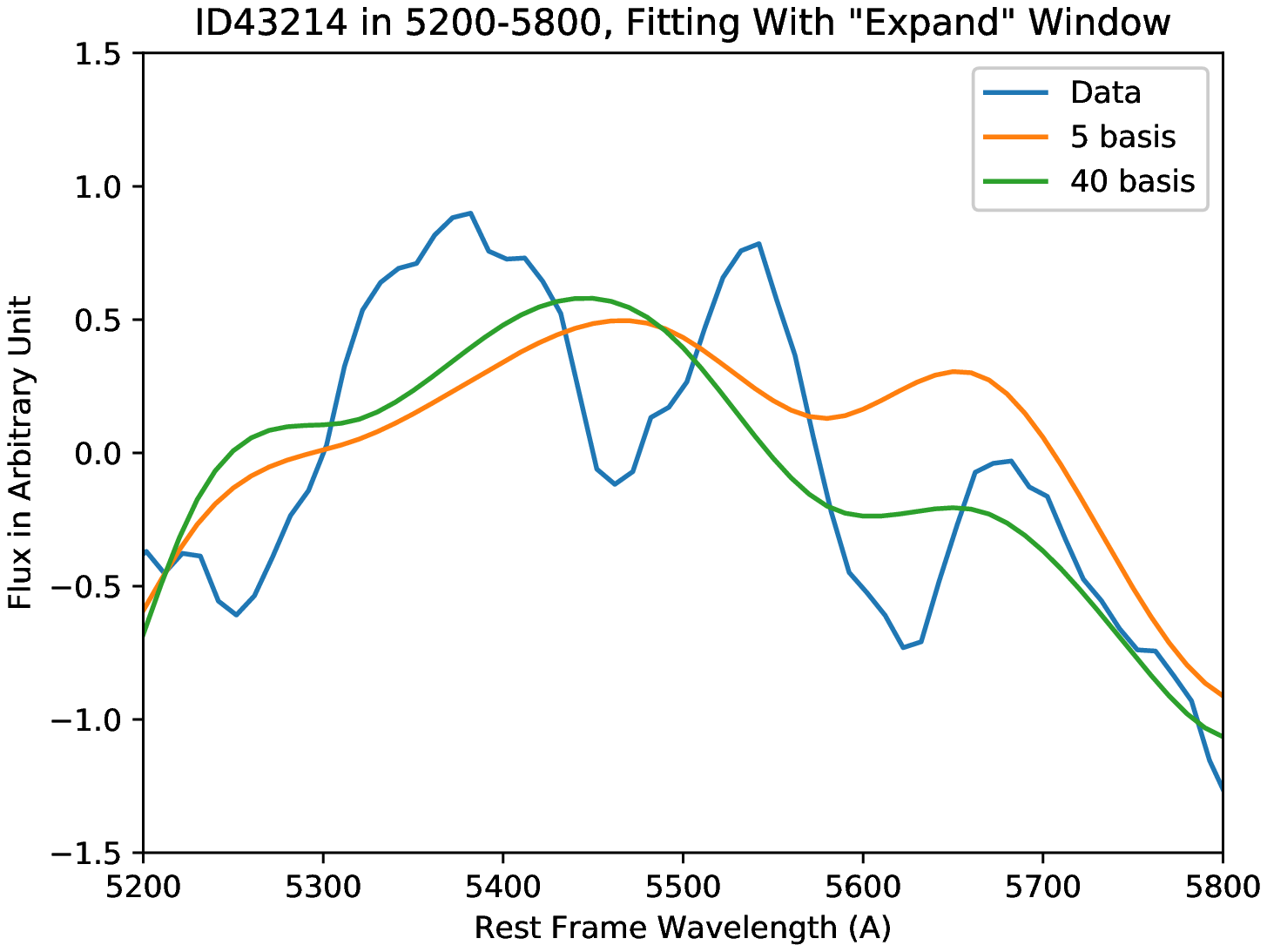}{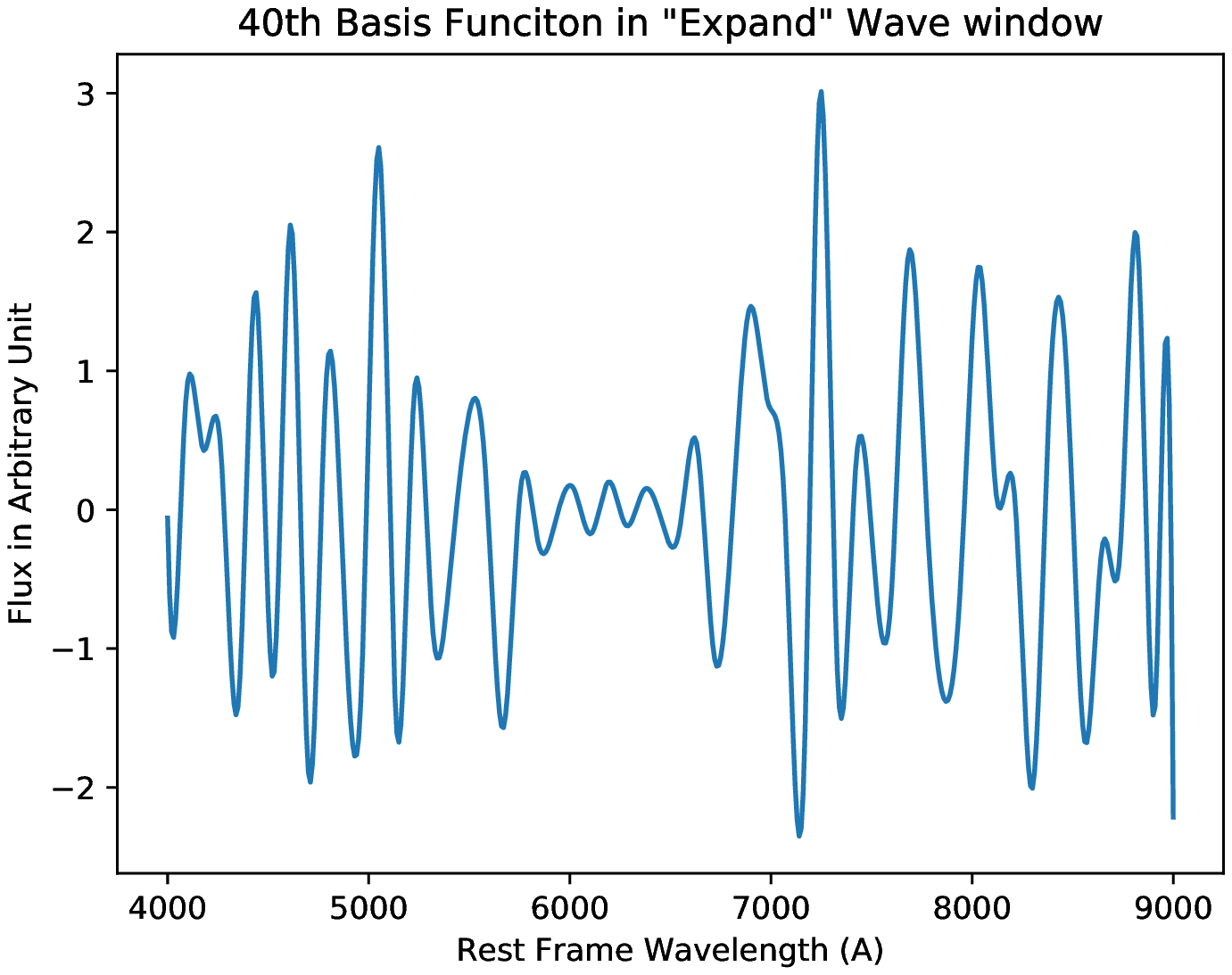}
\caption{Left: a close-up look of SN2017few spectrum between 5200 and 5800, the fitting result isn't desirable. The data here are compressed for the FPCA analysis. Right: the 40th basis function in `Expand' wave-window.}\label{fig:fitclose}
\end{figure}

We choose several pre-processed spectra to compare against the spectra re-constructed by FPCA algorithm, which is shown in Fig.\ref{fig:fitresult}. 
For SN2017few's spectrum here, its FPCA scores are $X(\lambda)\approx \mu(\lambda)-0.147\phi_1(\lambda)+0.669\phi_2(\lambda)+0.087\phi_3(\lambda)...$.  
For SN2013fr's spectrum here, its FPCA scores are$X(\lambda)\approx \mu(\lambda)-0.015\phi_1(\lambda)+0.282\phi_2(\lambda)+0.024\phi_3(\lambda)...$. 
The re-constructed spectra fits quite well when we choose 40 basis functions, however, some features are lost if we only choose 5 basis functions. 
Furthermore, we notice that the double-peak feature in SN2017few's spectrum at 5200-5800 \AA \ has not been reconstructed in the spectrum re-constructed by FPCA scores and basis functions, a close-up look is given in Fig.\ref{fig:fitclose}. 

\begin{figure}
\plottwo{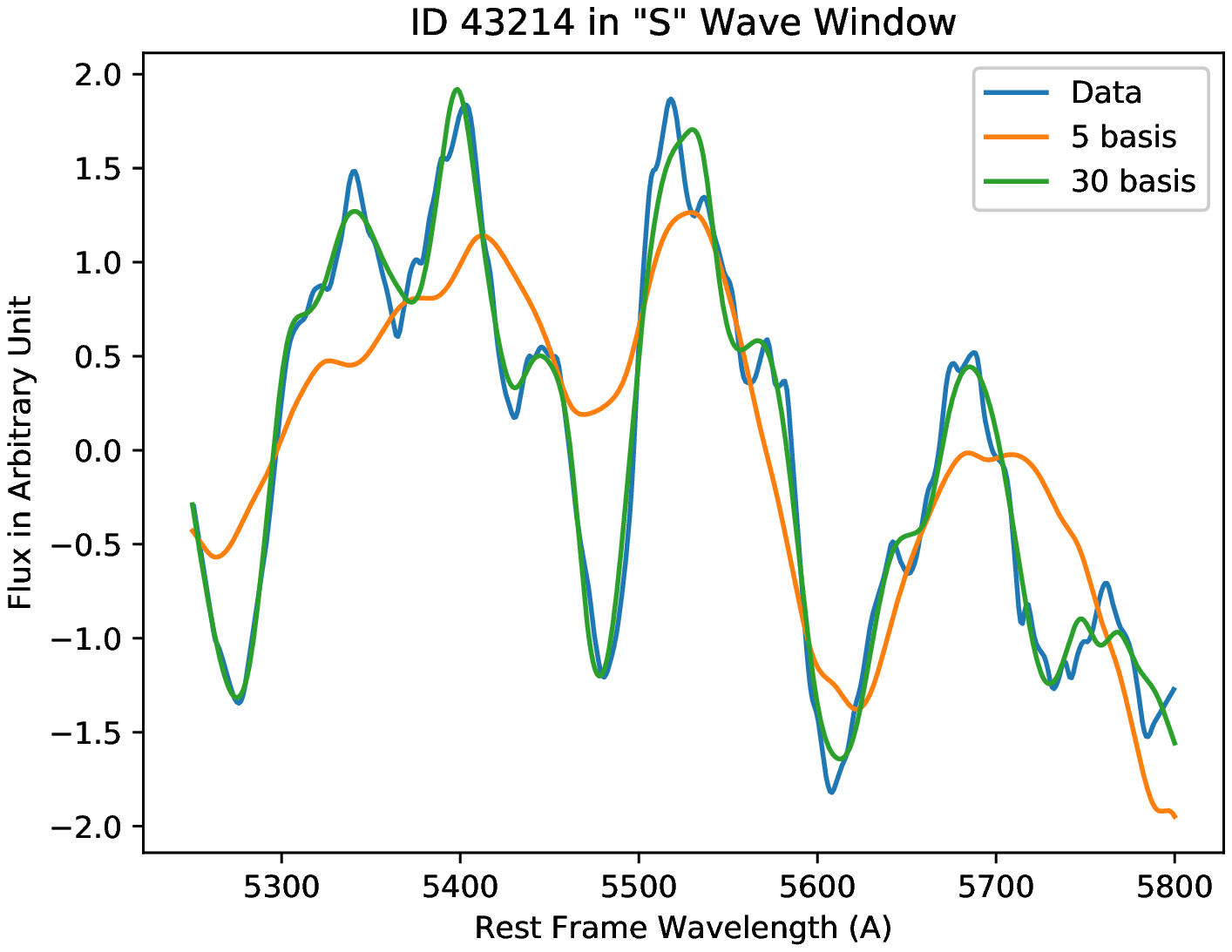}{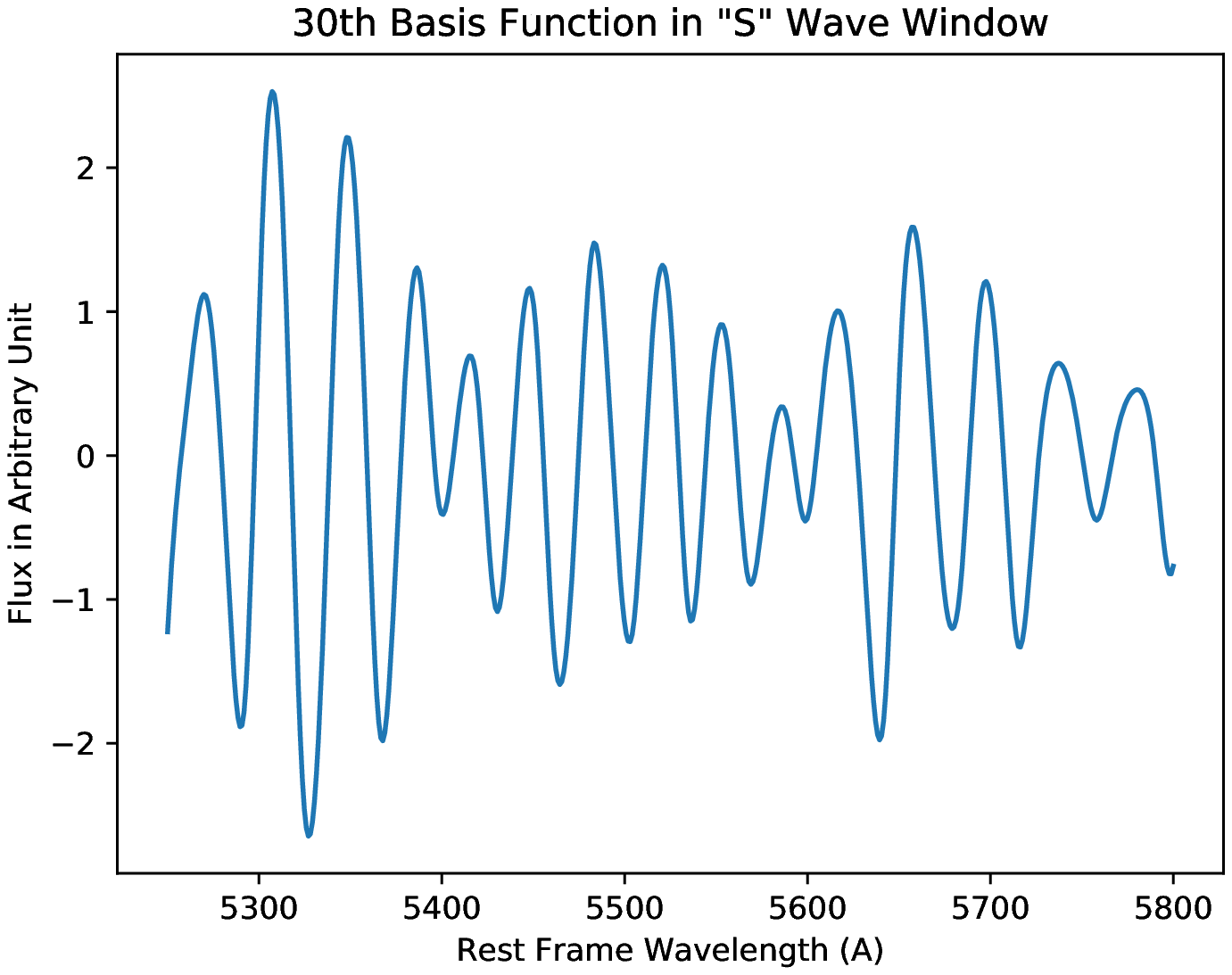}
\caption{Left: the fitting of SN2017few in `S' wave-window. In this scenario, the fitting fidelity is higher than `Expand' wave-window. To notice, because of the computational capacity, the spectra for FPCA in `Expand' window are down-sampled, which results in the data loss of the peak at 5350. Furthermore, the y-axis scale here differs from Fig.\ref{fig:fitclose} because the flux of different wave-windows are re-normalized seperately. Right: the 30th basis function in `S' wave-window.}\label{fig:goodfit}
\end{figure}

We choose the last (40th) basis function in `Expand' wave-window. 
Also, this basis function contains the most (24) peak among other `Expand' basis functions. 
The average distance between two peaks in the basis function is approximately 208 \AA, so we suggest that the spectral resolution of `Expand' wavewindow is 208 \AA. 
In contrast, the distance between the two peaks is approximately 200 \AA, which is similar to the wave resolution of `Expand' wave-window. 
If the relating spectra contains some extreamely sharp features like circumstellar material and the spectra of host galaxy, more basis functions are required to render the spectra. 
Alternatively, using multiple smaller wave-windows can avoid this problem. 
In the last (30th) basis function in `S' wave-window, there are 15 peaks and the resolution is 37 \AA, which indicates a better fitting result Fig.\ref{fig:goodfit}.

\clearpage

\section{Discussion about the relation of FPCA scores}\label{sec:twodimplot}

\begin{figure}[!htb]
\minipage{0.32\textwidth}
  \includegraphics[width=\linewidth]{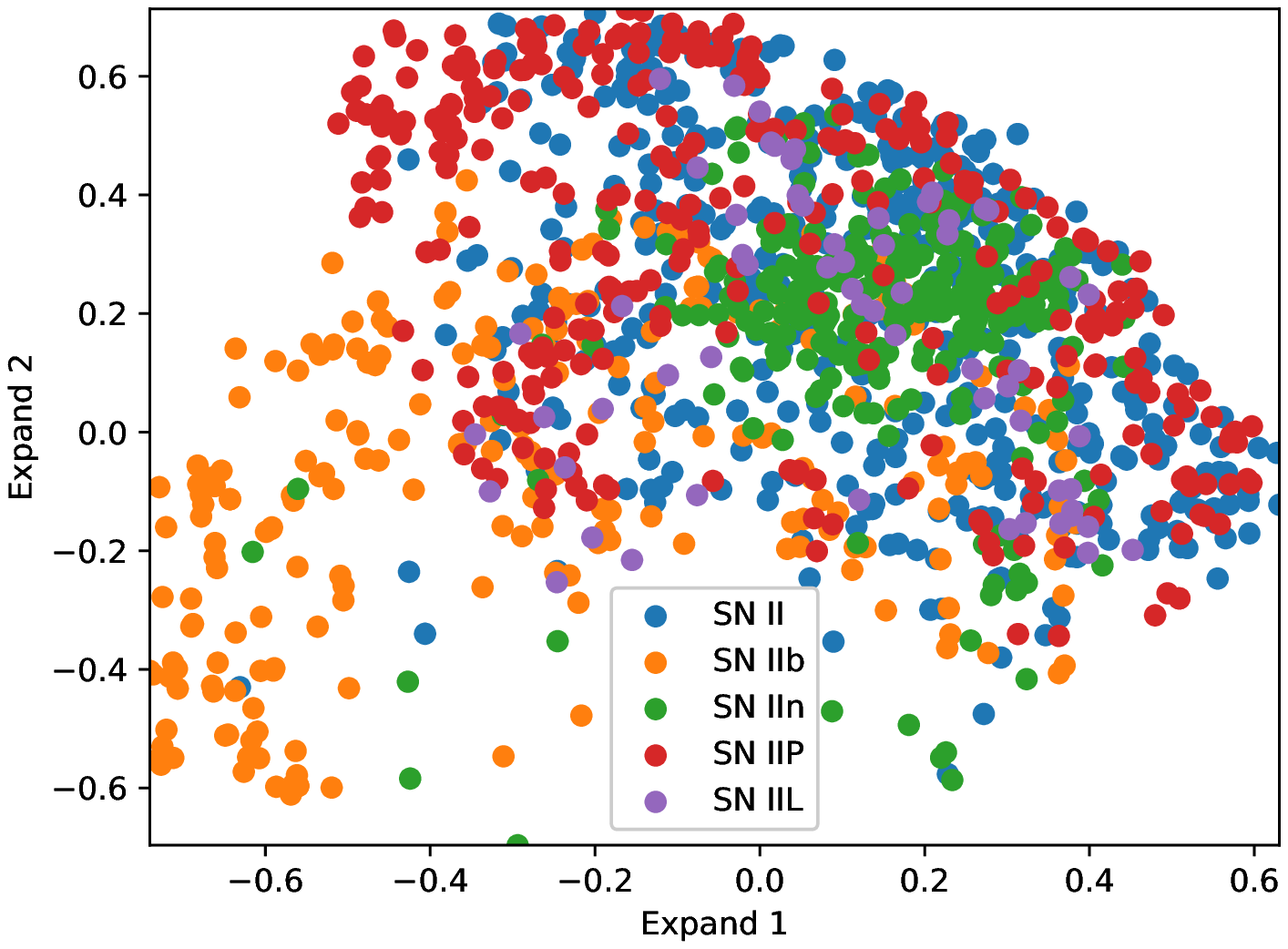}
  \caption{The relation between the first and second FPCA scores in `Expand' wave-window.}\label{fig:ex1andex2}
\endminipage\hfill
\minipage{0.32\textwidth}
  \includegraphics[width=\linewidth]{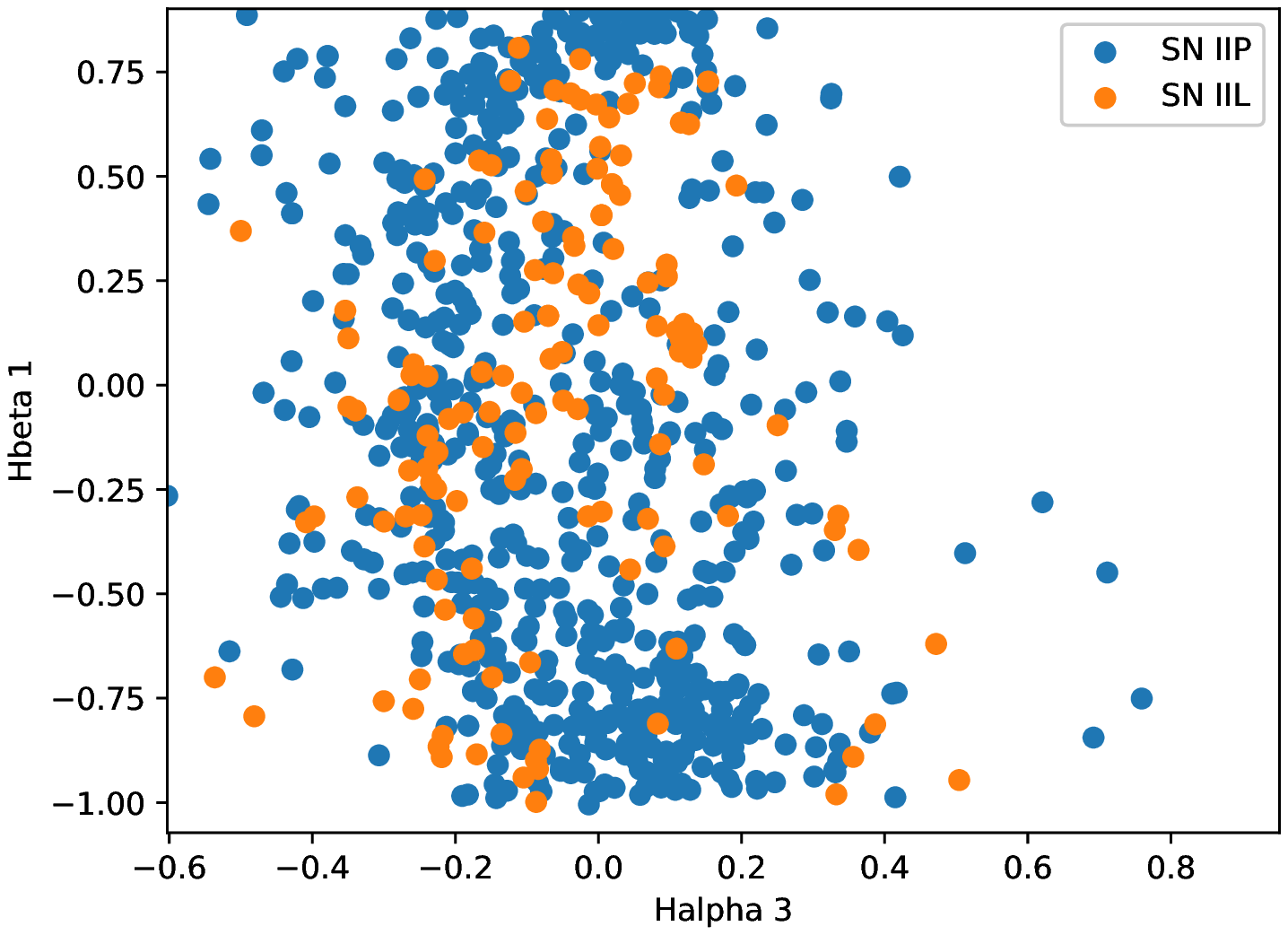}
  \caption{The relation between the `$H_\alpha$' and `$H_\beta$' 's third and first FPCA scores.}\label{fig:IIPLh1h3}
\endminipage\hfill
\minipage{0.32\textwidth}
  \includegraphics[width=\linewidth]{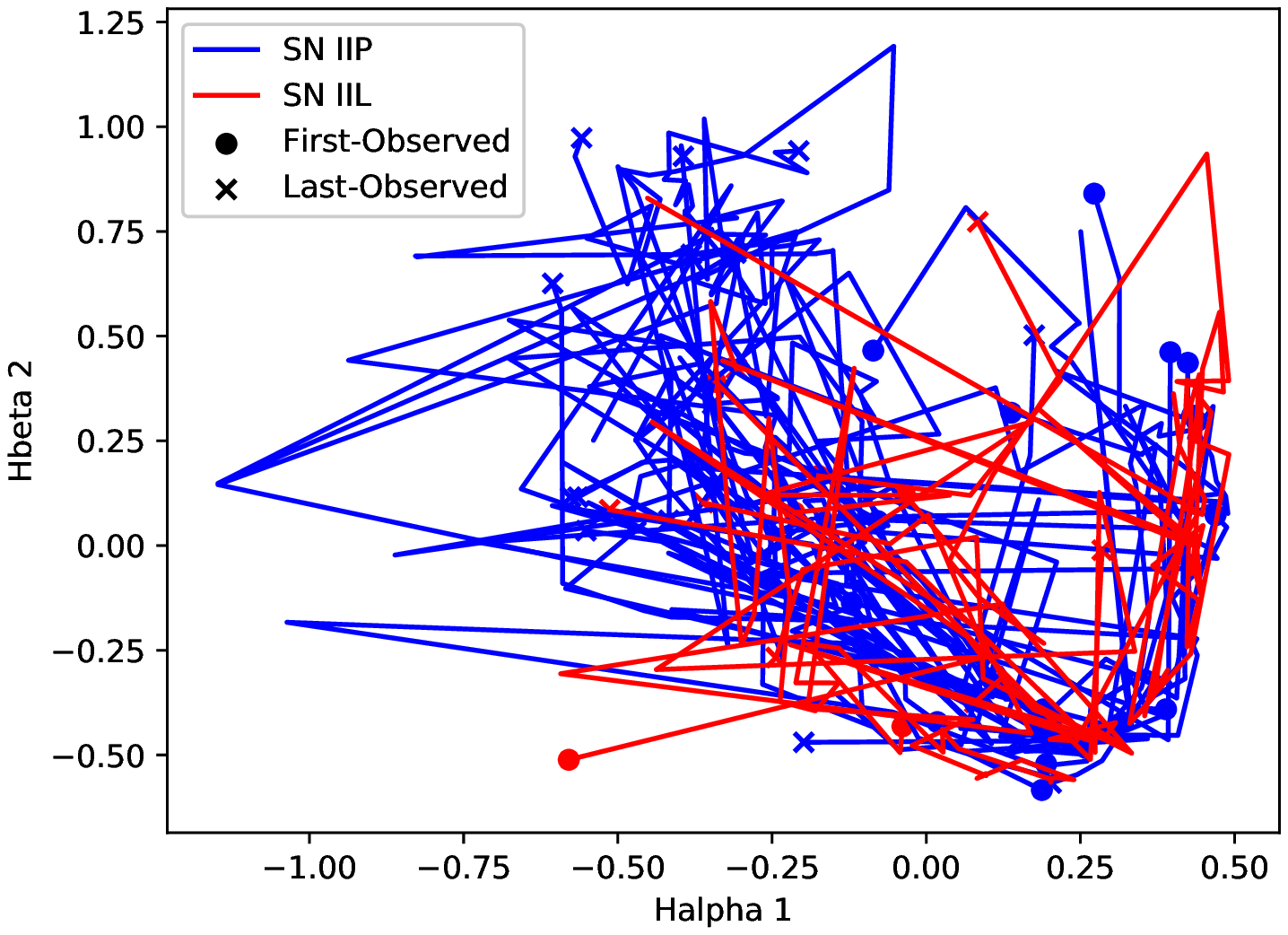}
  \caption{The evolution of Type IIP/IIL spectra, points are linked in the order of observation time.}\label{fig:timeevolution_two}
\endminipage
\label{fig:twodimplot}
\end{figure}

We plot some figure to illustrate the relation between two FPCA scores of different types of SNe. 
More plots are uploaded to the website. 
In Fig.\ref{fig:ex1andex2}, we notice Type IIb SNe are close to the lower-left side of the plot, while Type IIn SNe are clustered in the middle-upper side of the plot. For the other types of SNe, it is quite hard to distinguish them. 
A similar plot concentrated on Type IIP/IIL SNe is shown in Fig.\ref{fig:IIPLh1h3}, the dispersion of Type IIP and IIL SNe are not wholly overlapped, however, it is hard to simply seperate them with a curve. 
Plots using other FPCA scores stumbled upon same problem. 
We also discussed the evolution of spectra in the  FPCA scores' parametric space, Fig.\ref{fig:timeevolution} is a paradigm. 
No conspicuoous tendencies are observable in the plot. 
Because not all the SNe are intensively observed, most SNe are not well-observed in their whole lifetime, so we didn't take the phase of SNe' spectra into consideration. 

\clearpage

\section{The performance in high redshift spectra}

In this paper, no attempt for K-correction\citep{kcorrection} is done because most Type IIP/IIL objects' redshift are relatively low. 
However, we analyzed the performance of our classifiers on all 27 objects with redshifts higher than 0.04 \citep{kcorrection2}, all of them are Type IIP SNe. 
We use the cross validation method as is discussed in Section~\ref{sec:performance}, then count how many times the spectra is divided in the test set and how many times the spectra are misclassified. 
Considering the wavelength ranges of these spectra, we only choose `$H_\alpha$' wave window for trial. e

To sum up, the precision of high-redshift spectra in SVM is 0.610, the precision of high-redshift spectra in ANN is 0.627. 
The classifiers we have trained are not biased, the precision of Type IIP and Type IIL SNe are equal. 
Comparing to the overall precision of SVM (0.751) and ANN (0.851), we surmise the classifiers' performance are influenced by the redshift. 

\begin{deluxetable}{ccc|cc|cc}[ht!]
\tablecaption{High-z IIP Objects' Performance}
\tablehead{
\colhead{Id} & \colhead{Name} & \colhead{Redshift} & \colhead{Fail in SVM}  & \colhead{Test in SVM} & \colhead{Fail in ANN} & \colhead{Test in ANN}
}
\startdata
45175    &               SN2018aql &   0.074 &  33   &    95    &     0  &    11  \\
44559    &                SN2018pn &   0.048 &  10   &    91    &     0  &     6  \\
40509    &                SN2017hk &   0.059 &  89   &   109    &    16  &    18  \\
40495    &               SN2016ied &   0.0479&  101  &    106   &     6  &    10  \\
25600    &                OGLE15xg &   0.061 &  41   &    90    &     3  &    13  \\
21563    &        OGLE-2015-SN-052 &   0.06  &  62   &   81     &    13  &    15  \\
21454    &        OGLE-2015-SN-009 &   0.06  &   3   &   98     &     0  &     7  \\
20707    &                LSQ14gae &   0.041 &  12   &    83    &     2  &     5  \\
20196    &                LSQ14eeh &   0.09  &  92   &  111     &     1  &     6  \\
24058    &                LSQ13cuw &   0.25  &   8   &  113     &     0  &     6  \\
24059    &                LSQ13cuw &   0.25  &   1   &   78     &     2  &    12  \\
24060    &                LSQ13cuw &   0.25  &  81   &  104     &     5  &     8  \\
24057    &                LSQ13cuw &   0.25  &  44   &  101     &     5  &     9  \\
24055    &                LSQ13cuw &   0.25  &  53   &  100     &     2  &     5  \\
16147    &                LSQ13cuw &   0.25  &  37   &   98     &     4  &     6  \\
24050    &                LSQ13cuw &   0.25  &  18   &   87     &     6  &    15  \\
11399    &        OGLE-2013-SN-011 &   0.05  &  40   &   98     &     5  &     6  \\
19426    &        OGLE-2013-SN-011 &   0.05  &   1   &  105     &     0  &     9  \\
8755     & CSS120416-131835+213833 &   0.065 &  37   &    84    &     2  &     9  \\
8723     &                LSQ12blb &   0.052 &  95   &   107    &     3  &    12  \\
19049    &                LSQ12blb &   0.052 &  75   &    92    &     6  &    10  \\
27225    &                SN2011jj &   0.045 &   1   &   103    &     0  &    13  \\
41519    &                SN2007tu &   0.22  &  65   &   94     &     4  &    13  \\
41574    &                SN2007sx &   0.1171&   1   &    105   &     0  &    15  \\
42123    &             ESSENCEq060 &   0.1441&   7   &     86   &     4  &    14  \\
42122    &             ESSENCEq060 &   0.1441&  31   &    104   &     9  &    15  \\
42076    &             ESSENCEn271 &   0.241 &  33   &    88    &     3  &     7  \\
42110    &             ESSENCEm041 &   0.22  &   7   &   97     &     0  &    12  \\
42096    &             ESSENCEm003 &   0.219 &  24   &    83    &     7  &    12  \\
41502    &                SN2004fj &   0.1874&  26   &     94   &     4  &     8  \\
41513    &                SN2004fp &   0.2   &  16   &  81      &     2  &     8  \\
41568    &                SN2003lg &   0.2   &  15   &  93      &     3  &     6  \\
41380    &                SN2003kj &   0.0784&  68   &     88   &     6  &     9  \\
\enddata
\end{deluxetable}

\clearpage

\section{Table and Data}\label{sec:data_one}

\begin{deluxetable}{cc|ccc|ccc}[ht!]
\tablecaption{SVM performance in single wave-window}\label{tab:svmone}
\tablehead{
\colhead{Wave-window} & \colhead{C} & \colhead{Precision} & \colhead{Recall} & \colhead{F1-Score} 
& \colhead{$\sigma_P$} & \colhead{$\sigma_R$} & \colhead{$\sigma_F$}
}
\startdata
FeMg & 3020 & 0.797 & 0.796 & 0.793 & 0.050 & 0.050 & 0.050      \\
$H_\beta$ & 800000& 0.796 & 0.794 & 0.791 & 0.040 & 0.039 & 0.040  \\
FeOMgSi & 3020 & 0.775 & 0.774 & 0.771 & 0.049 & 0.049 & 0.050 \\
S & 3020 & 0.748 & 0.745 & 0.740 & 0.052 & 0.052 & 0.053       \\
Na & 80000000& 0.769 & 0.766 & 0.762 & 0.048 & 0.047 & 0.048     \\
$H_\alpha$ & 8000000 & 0.812 & 0.811 & 0.809 & 0.043 & 0.045 & 0.045 \\
Gap &800000000& 0.732 & 0.731 & 0.727 & 0.043 & 0.042 & 0.042 \\
NaMg &3020& 0.703 & 0.700 & {\bf 0.694} & 0.064 & 0.063 & 0.064     \\
Ca & 3020 & 0.793 & 0.791 & 0.787 & 0.059 & 0.059 & 0.060      \\
Visible & 3020 & 0.834 & 0.833 & {\bf 0.830} & 0.056 & 0.056 & 0.057 \\
Expand & 3020 & 0.822 & 0.820 & {\bf 0.815} & 0.079 & 0.079 & 0.081 \\
\enddata
\end{deluxetable}

\begin{deluxetable}{cc|ccc|ccc}[ht!]
\tablecaption{ANN performance in single wave-window}\label{tab:neuro}
\tablehead{
\colhead{Wave-window} &\colhead{Nodes in hidden layer} & \colhead{Precision} & \colhead{Recall} & \colhead{F1-Score} 
& \colhead{$\sigma_P$} & \colhead{$\sigma_R$} & \colhead{$\sigma_F$}
}
\startdata
FeMg & 90 & 0.810 & 0.809 & 0.807 & 0.051 & 0.052 & 0.053      \\
$H_\beta$ & 90 & 0.826 & 0.825 & 0.822 & 0.048 & 0.048 & 0.049  \\
FeOMgSi & 90 & 0.795 & 0.797 & 0.793 & 0.034 & 0.033 & 0.034 \\
S & 90 & 0.782 & 0.780 & 0.777 & 0.057 & 0.055 & 0.055       \\
Na & 40 & 0.806 & 0.804 & 0.799 & 0.051 & 0.049 & 0.051     \\
$H_\alpha$ & 90 & 0.851 & 0.850 & {\bf 0.849} & 0.038 & 0.038 & 0.038 \\
Gap & 40 & 0.750 & 0.749 & 0.747 & 0.063 & 0.064 & 0.064 \\
NaMg & 90 & 0.725 & 0.721 & 0.716 & 0.071 & 0.068 & 0.071     \\
Ca & 90 & 0.814 & 0.812 & 0.809 & 0.052 & 0.051 & 0.052      \\
Visible & 15 & 0.873 & 0.871 &  0.870 & 0.048 & 0.049 & 0.048 \\
Expand & 15 & 0.857 & 0.858 & 0.852 & 0.058 & 0.060 & 0.059 \\
\enddata
\end{deluxetable}

\begin{deluxetable}{c|ccc|ccc}[ht!]
\tablecaption{Performance in Different Dimension {\rm FPCA} Using ANN and `$H_\alpha$' Wave-Window}\label{tab:lowdimann_one}
\tablehead{
\colhead{Basis Functions}& \colhead{Precision} & \colhead{Recall} & \colhead{F1-Score} 
& \colhead{$\sigma_P$} & \colhead{$\sigma_R$} & \colhead{$\sigma_F$}
}

\startdata
30 & 0.851  &	0.85    &0.849	 & 0.038    &0.038  &	0.038  \\
25 & 0.847  &	0.848   & 0.846	 & 0.043    &0.041  &	0.043  \\
20 & 0.849  &	0.85    &0.847	 & 0.042    &0.045  &	0.046  \\
15 & 0.844  &	0.843   & 0.841	 & 0.035    &0.035  &	0.034  \\
10 & 0.842  &	0.842   & 0.839	 & 0.028    &0.026  &	0.028  \\
9  & 0.832  &	0.833   & 0.83	 & 0.038    &0.036  &	0.038  \\
8  & 0.821  &	0.818   & 0.816	 & 0.052    &0.053  &	0.054  \\
7  & 0.769  &	0.77    &0.767	 & 0.055    &0.057  &	0.056  \\
6  & 0.786  &	0.784   & 0.782	 & 0.049    &0.05 	& 0.05     \\
5  & 0.726  &	0.726   & 0.723	 & 0.038    &0.038  &	0.038  \\
4  & 0.696  &	0.694   & 0.688	 & 0.045    &0.044  &	0.044  \\
3  & 0.67 	&   0.67    &0.67	 & 0.059    &0.057  &	0.058  \\
2  & 0.658  &   0.656   & 0.65	 & 0.045    &0.043  &	0.045  \\
\enddata
\end{deluxetable}

\begin{deluxetable}{c|ccc|ccc}[ht!]
\tablecaption{Performance in Different Dimension {\rm FPCA} Using ANN and `Visible' Wave-Window}\label{tab:lowdimann}
\tablehead{
\colhead{Basis Functions}& \colhead{Precision} & \colhead{Recall} & \colhead{F1-Score} 
& \colhead{$\sigma_P$} & \colhead{$\sigma_R$} & \colhead{$\sigma_F$}
}

\startdata
50	&0.873	&0.871&	0.87	&0.048&  0.049 &	0.048    \\
40	&0.854	&0.854&	0.85	&0.038&  0.037 &	0.04     \\
35	&0.861	&0.859&	0.857	&0.053&  0.051 &	0.054    \\
30	&0.869	&0.869&	0.864	&0.042&  0.042 &	0.042    \\
25	&0.873	&0.876&	0.872	&0.049&  0.052 &	0.05     \\
20	&0.856	&0.854&	0.851	&0.051&  0.056 &	0.055    \\
15	&0.843	&0.845&	0.841	&0.051&  0.051 &	0.052    \\
10	&0.85	&0.845&	0.842	&0.079&  0.079 &	0.083    \\
8	&0.872	&0.873&	0.87	&0.073&  0.073 &	0.073    \\
7	&0.824	&0.822&	0.819	&0.066&  0.065 &	0.064    \\
6	&0.775	&0.777&	0.774	&0.05&   0.05 	& 0.048      \\
4	&0.761	&0.763&	0.755	&0.054&  0.055 &	0.053    \\
2	&0.707	&0.702&	0.7    & 0.078&  0.079 &   0.079     \\
\enddata
\end{deluxetable}

\end{document}